\documentclass[a4paper, 12pt, twoside]{article}
\usepackage{graphicx}
\usepackage{epsfig}
\usepackage{subfigure}
\usepackage{graphics}
\usepackage{amsmath}
\usepackage{amsfonts}
\usepackage{amssymb}
\usepackage{url}
\usepackage{hyperref}

\begin{document}
  \graphicspath{{images/}}
  \newcommand{\kev}{{\,\text{ke}\eVdist\text{V\/}}}
\newcommand{\mev}{{\,\text{Me}\eVdist\text{V\/}}}
\newcommand{\gev}{{\,\text{Ge}\eVdist\text{V\/}}}
\newcommand{\gevc}{{\,\text{Ge}\eVdist\text{V\//\ensuremath{c}}}}
\newcommand{\gevcc}{{\,\text{Ge}\eVdist\text{V\//\ensuremath{c^{2}}}}}
\newcommand{\tev}{{\,\text{Te}\eVdist\text{V\/}}}
\newcommand{\pbi}{\,\text{pb}\ensuremath{^{-1}}}
\newcommand{\fbi}{\,\text{fb}\ensuremath{^{-1}}}
\newcommand {\ptmiss}{\mbox{$\not\hspace{-0.55ex}{P}_t$}}
\newcommand {\misspt}{\mbox{$\not\hspace{-0.55ex}{P}_t$}}
\newcommand{\Cpp}{C\raisebox{2pt}{\scriptsize{++}}}
\newcommand{\sol}{\ensuremath{S_{0}^{L}}}
\newcommand{\sor}{\ensuremath{S_{0}^{R}}}
\newcommand{\tsor}{\ensuremath{\widetilde{S}_{0}^{R}}}
\newcommand{\sil}{\ensuremath{S_{1}^{L}}}
\newcommand{\viiil}{\ensuremath{V_{1/2}^{L}}}
\newcommand{\viiir}{\ensuremath{V_{1/2}^{R}}}
\newcommand{\tviiil}{\ensuremath{\widetilde{V}_{1/2}^{L}}}
\newcommand{\vol}{\ensuremath{V_{0}^{L}}}
\newcommand{\vor}{\ensuremath{V_{0}^{R}}}
\newcommand{\tvor}{\ensuremath{\widetilde{V}_{0}^{R}}}
\newcommand{\vil}{\ensuremath{V_{1}^{L}}}
\newcommand{\siiil}{\ensuremath{S_{1/2}^{L}}}
\newcommand{\siiir}{\ensuremath{S_{1/2}^{R}}}
\newcommand{\tsiiil}{\ensuremath{\widetilde{S}_{1/2}^{L}}}
\newcommand{\be}{\begin{equation}}
\newcommand{\ee}{\end{equation}}
\newcommand{\ud}{\mathrm{d}}
\newcommand{\gimmunnu}{g^{\mu \nu}}
\newcommand{\pimmu}{p^{\mu}}
\newcommand{\pinnu}{p^{\nu}}
\newcommand{\qunnu}{q^{\nu}}
\newcommand{\qummu}{q^{\mu}}
\newcommand{\Tab}[1]{Table~\ref{tab:#1}}
\newcommand{\Tabs}[2]{Tables~\ref{tab:#1} and~\ref{tab:#2}}
\newcommand{\Sec}[1]{Sec.~\ref{sec:#1}}
\newcommand{\Secs}[2]{Secs.~\ref{sec:#1} and~\ref{sec:#2}}
\newcommand{\App}[1]{Appendix~\ref{sec:#1}}
\newcommand{\Eq}[1]{Eq.~(\ref{eq:#1})}
\newcommand{\Eqs}[2]{Eqs.~(\ref{eq:#1}) and (\ref{eq:#2})}
\newcommand{\Fig}[1]{\mbox{Fig.~\ref{fig:#1}}}
\newcommand{\Figs}[2]{Figs.~\ref{fig:#1} and~\ref{fig:#2}}
\newcommand{\FIG}[2][z]{\mbox{Figure~\ref{fig:#2}\ifthenelse{\equal{#1}{z}}{}{(#1)}}}
\newcommand{\Chap}[1]{Ch.~\ref{ch:#1}}
\newcommand{\spazio}{\rule{0pt}{15pt}}
\newcommand{\spaziosmall}{\rule{0pt}{4pt}}
\newcommand{\mycaption}[1]{\begin{minipage}{0.9\textwidth}\caption{\small{#1}}\end{minipage}}
\newcommand{\mycaptionshort}[2]{\caption[\small{#1}]{\small{#2}}}
\newcommand{\rnge}{\hbox{$\,\text{--}\,$}}
\newcommand{\cm}{\,\text{cm}}
\newcommand{\DO}{{D{\O}}\xspace}
\newcommand{\cspeed}{\ensuremath{c}}
\newcommand{\alphas}{\ensuremath{\alpha_{S}}}
\newcommand{\lambdaqcd}{\ensuremath{\Lambda_{QCD}}}
\newcommand{\diff}[1]{\ensuremath{\mathrm{d}}#1}
\newcommand{\pt}{\ensuremath{p_{T}}}
\newcommand{\kt}{\ensuremath{k_{\perp}}}
\newcommand{\qcut}{\ensuremath{q_{cut}}}
%%%%%particles
\newcommand{\Z}{\ensuremath{Z}}
\newcommand{\photon}{\ensuremath{\gamma}}
\newcommand{\Wpm}{\ensuremath{W^{\pm{}}}}
\newcommand{\Wp}{\ensuremath{W^{+}}}
\newcommand{\Wm}{\ensuremath{W^{-}}}
\newcommand{\gluon}{\ensuremath{g}}
\newcommand{\zjets}{\ensuremath{Z/\gamma^{*}+jets}}

%%%%%%%%%%%%%%%
%%%%%UNITS
\newcommand{\mum}{\ensuremath{\rm{\mu}}\text{m}}
\newcommand{\second}{\text{s}}
%\newcommand{\fbi}{\,\text{fb}^{-1}}
%%%%%Generator names
\newcommand{\genname}[1]{\texttt{\textbf{#1}}}
\newcommand{\pythia}{\genname{PYTHIA}}
\newcommand{\alpgen}{\genname{AlpGen}}
\newcommand{\sherpa}{\genname{SHERPA}}
\newcommand{\herwig}{\genname{HERWIG}}
\newcommand{\madgraph}{\genname{MADGRAPH/MADEVENT}}
\newcommand{\mcatnlo}{\genname{MC@NLO}}
\newcommand{\amegic}{\genname{AMEGIC++}}
\newcommand{\apacic}{\genname{APACIC++}}
%%%%%%%%%%%%%%%%%%%%%%%%%%%%%%%%%%%  
\newcommand{\MYANNOT}[1]{%
%%% commenta questo per segare le MYANNOT
\fbox{\begin{minipage}{3cm}{\tiny{=$\widehat{\rule{2mm}{0pt}}$.$\widehat{\rule{2mm}{0pt}}$= #1}}\end{minipage}}
}%%%%%%%%%%%%%%%%%%%%%%%%%%%%%%%%%%% 

  \begin{titlepage}
  \pagestyle{plain}
  %%%%% PREPRINT NUMBERS %%%%%%
  \begin{flushright}
  MCnet/09/06\\
  \end{flushright}
  %%%%%%%%%%%%%%%%%%%%%%%%%%%%%%
  \vspace{4\baselineskip}
  %%%%%%%%%%%%%%%%%%% TITLE %%%%%%%%%%%%%%%%%%
  \begin{center}
    {\Large\bf 
      A study on Matrix Element corrections\\
      in inclusive $Z/\gamma^{*}$ production at LHC\\
      as implemented in\\
      \pythia, \herwig, \alpgen\ and \sherpa.
    }
  \end{center}
  %%%%%%%%%%%%%%%% AUTHORS %%%%%%%%%%%%%%%%%%%%%%%
  %\vspace{1cm}
  \begin{center}
    P. Lenzi\footnote{\href{mailto:piergiulio.lenzi@cern.ch}{piergiulio.lenzi@cern.ch}} \\Dept. of Physics, Universit\`a degli Studi di Firenze

    \vspace{0.5cm}

    J. M. Butterworth \\Dept. of Physics and Astronomy, University College London
  \end{center}
  %%%%%%%%%%%%%%%%%%%%%%% AFFILIATION %%%%%%%%%%%%
  %\vskip 10mm
  %\setlength{\evensidemargin}{\oddsidemargin}
\begin{abstract}
We study Matrix Element corrections as implemented in four popular event generators for hadron collisions. We compare \pythia, \herwig, \alpgen\ and \sherpa\ in the $Z/\gamma^{*}$ inclusive production at LHC.
\pythia\ and \herwig\ are able to correct the first emission from the shower taking the Matrix Element calculation for one additional parton into account.
\sherpa\ and \alpgen\ are able to take into account Matrix Element corrections not only for one, but rather for several hard emissions from the incoming partons. This can be done at the price of introducing a separation cut to distinguish a Matrix Element and a Parton Shower populated regions.
In this paper we check the effect of Matrix Element corrections in \pythia\ and \herwig\ and we check that results from these two generators are consistent. 
Then we turn to \sherpa\ and \alpgen, that implement two different methods to match Matrix Element calculations and Parton Shower. 
If we constraint them so that no more than one parton can emerge from the Matrix Element calculations they should both give results similar to \pythia\ and \herwig.
In other words \pythia\ and \herwig\ provide us with the correct reference to spot possible issues with the matching prescriptions implemented in \sherpa\ and \alpgen.
We also check to what extent the dependency on the Matrix Element - Parton Shower separation cut is canceled in these two generators.

\end{abstract}
\end{titlepage}

\section{Introduction}
The description of the QCD radiation pattern that accompanies the partons involved in a hard scattering process can be done at different precision levels. 
Most event generators attach the QCD radiation to the partons using the Parton Shower technique. 
While this approach provides a good description of many low \pt\ observables it may fail in efficiently filling the phase space for hard radiation. 
This limitation is connected to the Parton Shower being a collinear approximation of the decription of parton splittings.

A way to improve the description of the QCD radiation pattern is to augment the Parton Shower with information coming from the exact matrix element calculation.
The generators studied in this work follow this approach, but with significant differences.
\pythia~\cite{Sjostrand:2006za} and \herwig~\cite{Corcella:2002jc,Bahr:2008pv} are able to modify the shower in such a way that the hardest emission is described using the exact Matrix Element calculation  for one additional real emission. 
\alpgen~\cite{Mangano:2002ea} and \sherpa~\cite{Gleisberg:2008ta} are able to take Matrix Element corrections for several (not only one) hard emissions. The main idea in both generators is that configurations in which the emitted partons' \pt\ is below a certain threshold are described with a pure Parton Shower approach, while configurations in which $n$ partons are above the threshold are described with the $n$-real emissions Matrix Element.
So, at the price of introducing an arbitrary threshold that is not present in \pythia\ or \herwig, \alpgen\ and \sherpa\ should be able to describe the emission of several hard partons with the corresponding Matrix Element calculation.
The dependency on the arbitrary threshold has to be as limited as possible.

In this paper we compare these different approaches for the case of the inclusive $Z/\gamma^{*}$ production at LHC.
We first check the effect of Matrix Element corrections in \pythia\ and \herwig. We discuss the differences in the implementation and compare results from these two generators.
Then, we discuss the differences in the merging prescriptions implemented in \sherpa\ and \alpgen\ and we present a sort of a consistency test about these prescriptions.
The main idea behind this test is that if we allow the Matrix Element generators in \sherpa\ and \alpgen\ to emit not more than one additional parton we should recover results from \herwig\ and \pythia, because we are using the same Matrix Element content.
We also check to what extent the dependency on the threshold value in \alpgen\ and \sherpa\ is canceled.

\section{Improving the description of the QCD\\radiation pattern using Matrix Element\\Calculations}
In this section we briefly describe the implementation of the Matrix Element corrections in the generators studied in this work. 
We group the \pythia\ and \herwig\ implementations under the name of ``Parton Shower reweighting'', while \sherpa\ and \alpgen\ belong to the cathegory of ``Matrix Element - Parton Shower merging''.

\subsection{Parton Shower reweighting}
In this section we describe the implementation of Matrix Element corrections in the Initial State Parton Shower as implemented in \pythia. We will enlight the differences between \herwig\ and \pythia\ at the end of this section.

\begin{figure}[!h]
  \centering\includegraphics[width=0.38\textwidth]{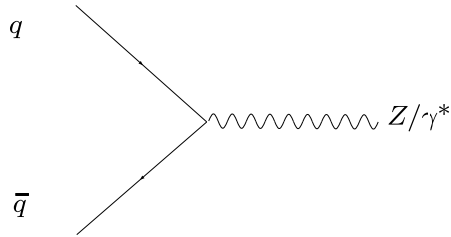}
  \caption{Lowest order contribution to the $Z/\gamma^{*}$ production.\label{fig:lowestorder}}
\end{figure}

\begin{figure}[!h]
  \centering
  \subfigure[]{\includegraphics[width=0.38\textwidth]{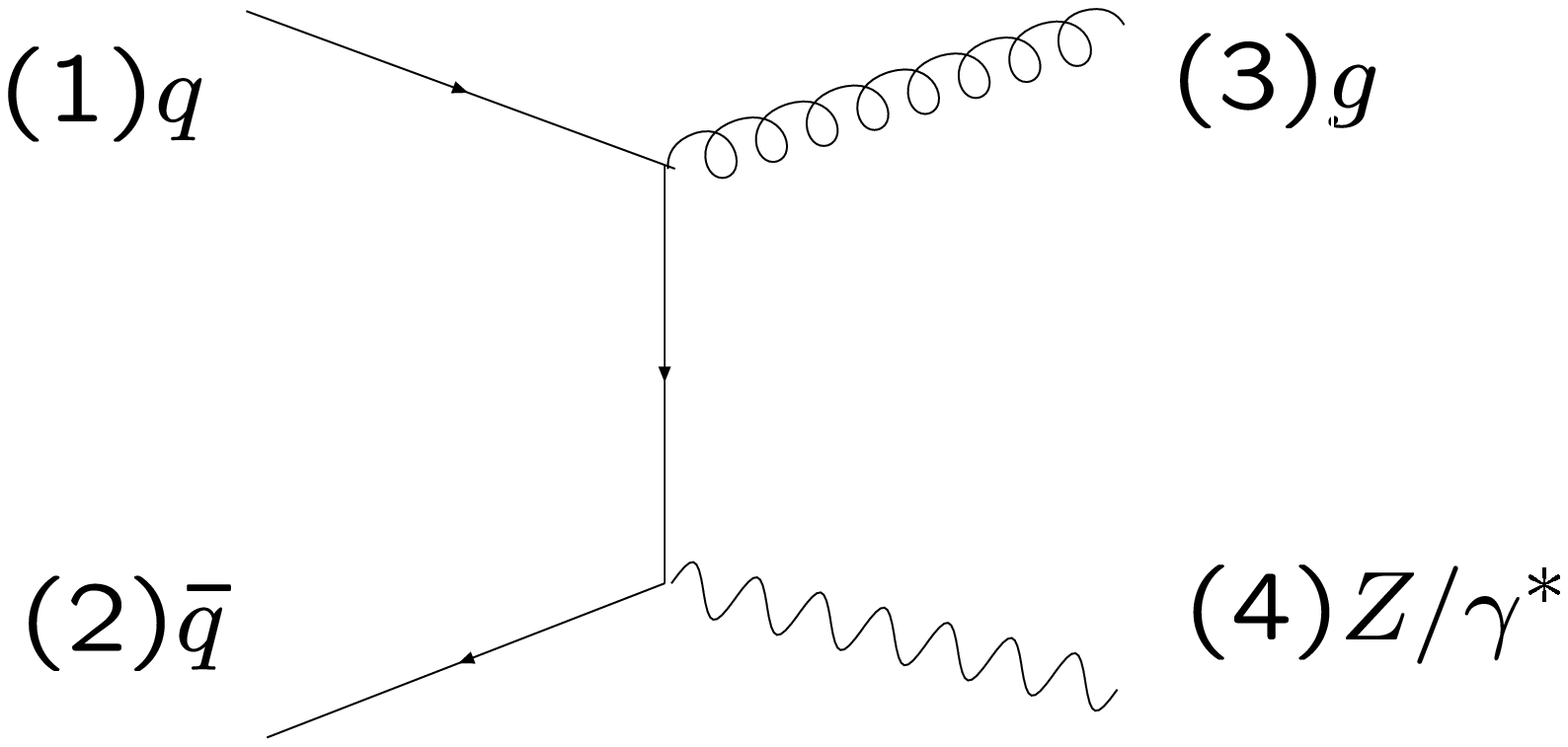}}
  \subfigure[]{\includegraphics[width=0.38\textwidth]{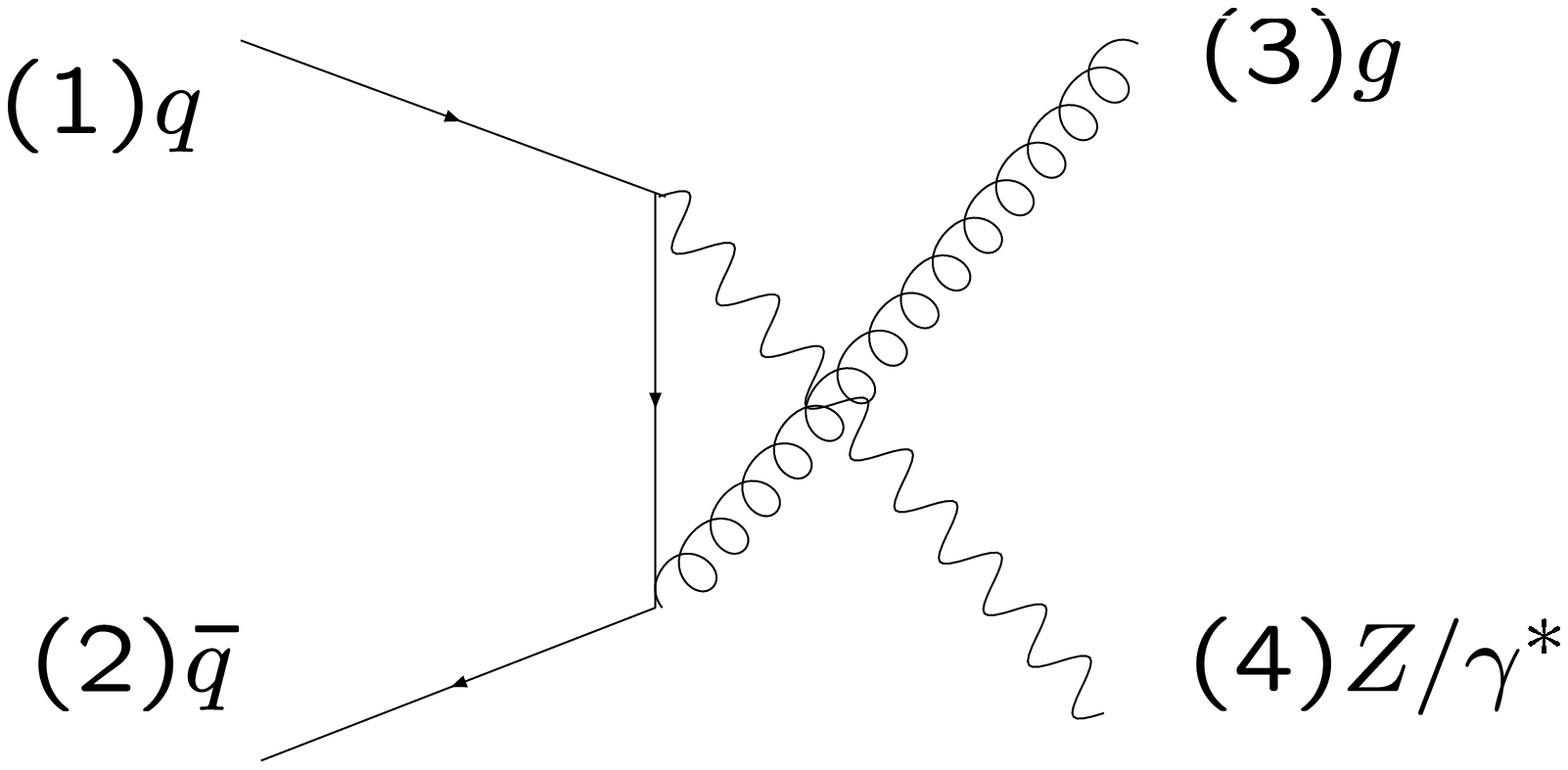}}
  \caption{Graphs contributing to the $q\bar{q}\rightarrow{}Z/\gamma^ {*}g$ process.\label{fig:qqzg}}
\end{figure}

\begin{figure}[!h]
  \centering
  \subfigure[]{\includegraphics[width=0.38\textwidth]{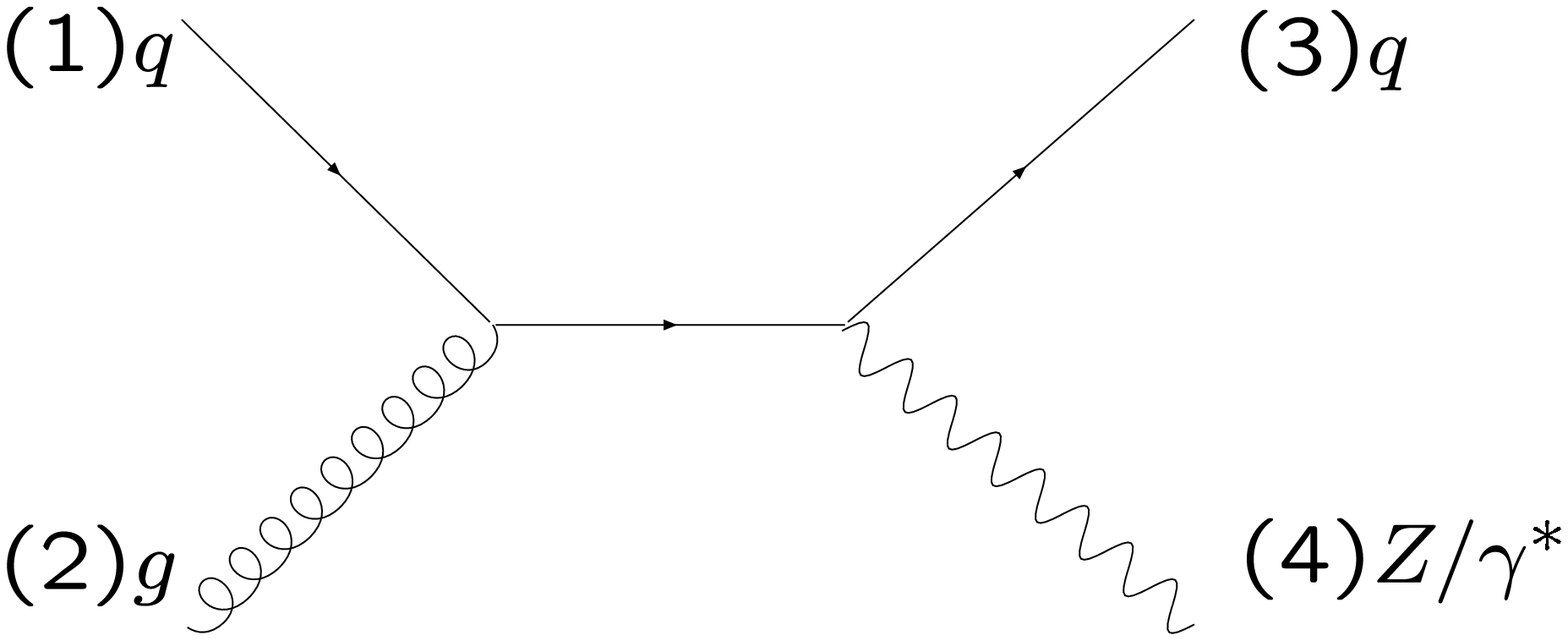}}
  \subfigure[]{\includegraphics[width=0.38\textwidth]{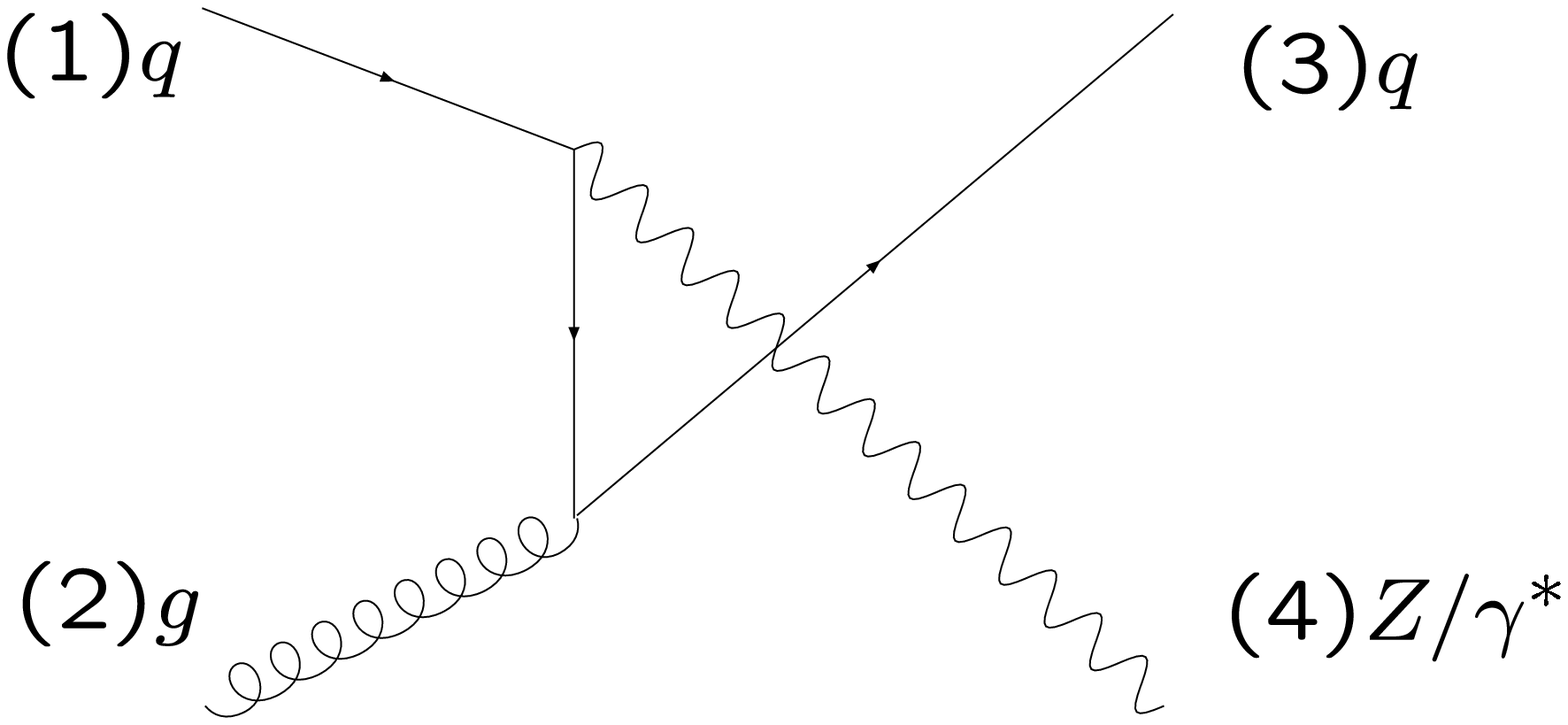}}
  \caption{Graphs contributing to the $qg\rightarrow{}Z/\gamma^ {*}q$.\label{fig:qgzq}}
\end{figure}

Matrix element corrections for inclusive $Z/\gamma^{*}$ production in \pythia\ modifies the first emission from the PS so that the effect of the first order correction is reproduced.
The lowest order graph contributing to $Z/\gamma^{*}$ production is shown in \Fig{lowestorder}.
The processes that contribute to the first order correction are $q\bar{q}\rightarrow{}Z/\gamma^{*}g$, represented by the graphs in \Fig{qqzg}, and $qg\rightarrow{}Z/\gamma^{*}q$, represented by the graphs in \Fig{qgzq}.

\begin{figure}[!h]
  \centering
  \includegraphics[width=0.45\textwidth]{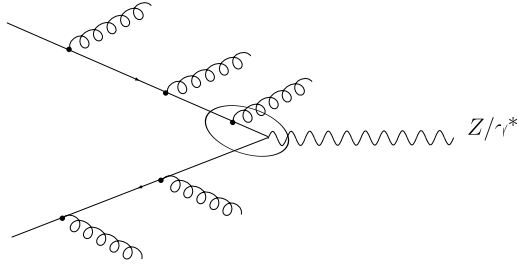}
  %\subfigure[]{\includegraphics[width=0.45\textwidth]{ps-qqzg-like-1.pdf}}
  %\subfigure[]{\includegraphics[width=0.45\textwidth]{ps-qqzg-like-2.pdf}}
  \caption{The circled PS branching is considered of type $q\bar{q}\rightarrow{}Z/\gamma^ {*}g$.\label{fig:qqzg-like}}
\end{figure}

\begin{figure}[!h]
  \centering
  \includegraphics[width=0.45\textwidth]{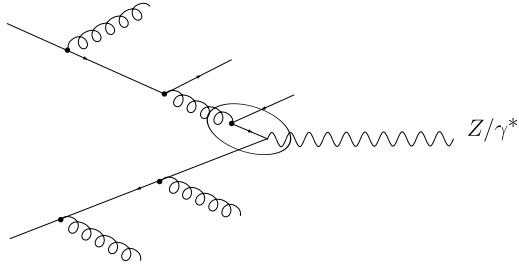}
  \caption{The circled PS branching is considered of type $qg\rightarrow{}Z/\gamma^ {*}q$.\label{fig:qgzq-like}}
\end{figure}

The correction comes as a re-weighting factor of the parton shower. The initial state parton shower is weighted with two factors, one to reproduce the matrix elements of \Fig{qqzg}, $W_{q\bar{q}\rightarrow{}Z/\gamma^{*}g}$, and one to reproduce the matrix element of \Fig{qgzq}, $W_{qg\rightarrow{}Z/\gamma^{*}q}$ \cite{Miu:1998ju}.
%\Fig{qgzq} represent the graphs contributing to the first order correction $qg\rightarrow{}Z/\gamma^{*}q$.
The PS emission closest to the hard $q\bar{q}$ process is the one that gets the correction.
In order to correct the PS we need to classify this emission as either $(q\bar{q}\rightarrow{}Z/\gamma^ {*}g)$-like or $(qg\rightarrow{}Z/\gamma^ {*}q)$-like.
A PS branching like the one depicted in \Fig{qqzg-like} is considered $(q\bar{q}\rightarrow{}Z/\gamma^ {*}g)$-like, while a branching like the one in \Fig{qgzq-like} is considered $(qg\rightarrow{}Z/\gamma^ {*}q)$-like.

The ratio of the ME and PS differential cross sections, classified as described above, leads to the following expressions for $W_{q\bar{q}\rightarrow{}Z/\gamma^{*}g}$ and $W_{qg\rightarrow{}Z/\gamma^{*}q}$ \cite{Miu:1998ju}:
\begin{eqnarray}
  W_{q\bar{q}\rightarrow{}Z/\gamma^{*}g}=\frac{\hat{t}^{2}+\hat{u}^{2}+2m_{Z}^{2}\hat{s}}{\hat{s}^{2}+m_{Z}^{4}},\\
  W_{qg\rightarrow{}Z/\gamma^{*}q}=\frac{\hat{s}^{2}+\hat{u}^{2}+2m_{Z}^{2}\hat{t}}{(\hat{s}-m_{Z}^{2})^{2}+m_{Z}^{4}},
\end{eqnarray}
where $\hat{s}$, $\hat{t}$ and $\hat{u}$ are the Mandelstam variables.
It can be shown that 
\begin{equation}
  \frac{1}{2}<W_{q\bar{q}\rightarrow{}Z/\gamma^{*}g}<1,
  \label{eq:wqqbarzg}
\end{equation}
and
\begin{equation}
  1<W_{qg\rightarrow{}Z/\gamma^{*}q}<3.
  \label{eq:wqgzq}
\end{equation}
This means that the Parton Shower alone would overestimate the $q\bar{q}\rightarrow{}Z/\gamma^{*}g$ contribution and underestimate the $qg\rightarrow{}Z/\gamma^{*}q$ contribution.

As described in \cite{Miu:1998ju} the correction is performed in two steps: first the starting scale for the shower is raised to the hadronic center of mass energy, so that any hard emission from the shower is kinematically possible; then the first splitting from the shower is classified as $(q\bar{q}\rightarrow{}Z/\gamma^ {*}g)$-like or $(qg\rightarrow{}Z/\gamma^ {*}q)$-like and the splitting probability is modified according to the corresponding $W$ function.

Matrix element corrections are implemented in \herwig\ in a similar way, but with two important differences:
\begin{itemize}
  \item in \herwig\ the PS is angular ordered, which means that the early emissions are soft large angle gluons; thus, the emission that deserves the corrections is not the first like in \pythia, as pointed in \cite{Seymour:1994df};  
  \item in \herwig\ the PS cannot fill the phase space for values above the hard scale of the process; in this respect it is similar to \pythia\ with starting scale set to $M_{Z}$.
\end{itemize}
For these reasons a two-step approach is used in \herwig\ to implement ME corrections \cite{Corcella:1999gs}.
In the phase space region covered by the PS, corrections are applied as in \pythia, the only difference being that not the first emission, but rather the \emph{hardest emission so far} during the shower evolution gets the correction \cite{Seymour:1994df}. These corrections are referred to as ``soft ME corrections''.
In the ``dead zones'', that are left completely uncovered by the PS, the exact ME for one additional parton is used, with subsequent PS. These corrections are referred to as ``hard ME corrections''.

\subsection{Matrix Element - Parton Shower merging}
\label{sec:ckkwmlm}
\sherpa\ and \alpgen\ implement matrix element corrections in a different ways. They both subdivide the phase space in two regions, one for jet production which is filled by the Matrix Element calculation, and one for the jet evolution, which is filled by the Parton Shower.
The criterion that is used to distinguish the two regions is a jet measure based cut.

The procedure implemented in \sherpa\ is known as CKKW merging. It was first proposed for lepton collisions in \cite{Catani:2001cc}, then it was extended to hadron collisions in \cite{Krauss:2002up}.
In a nutshell, the prescription foresees a preliminary step in which cross sections are calculated for up to $n$ additional partons in the final state; ME level events are produced according to the calculated cross sections and they are weighted with a Sudakov weight that makes these states exclusive. 
In other words the Sudakov weight accounts for the probability that, given an $n$ parton final state, no further hard emission is done by the Parton Shower; the reason for this is that any additional hard emission is going to be simulated by the $n+1$ Matrix Element.

More extensively, the whole procedure foresees the following steps:
\begin{enumerate}
  \item ME cross sections $\sigma_{n,i}$ are calculated for each parton multiplicity $n$ and for each different combination $i$ of partons that contributes to multiplicity $n$. A cutoff $y_{cut}$ on the separation of partons is applied to avoid divergences.
  A fixed \alphas$^{ME}$ is used.
  \item One among configurations $n,i$ is selected with probability $P_{n,i}=\sigma_{n,i}/\sum_{m,j}\sigma_{m,j}$.
  \item Parton momenta are generated according to the corresponding matrix element squared.
  \item The scales at which the splittings happened are reconstructed: this is achieved through a \kt\ clustering of the partons emerging from the ME. The clustering is stopped when the core $2\rightarrow{}2$ process is found.
  This leads to a series of $n-2$ clusterings with associated values of the \kt\ distance $y_{2}...y_{n}$. 
  Once the values $y_{i}=t_{i}^{2}/s$ are known we can finally calculate the ME event weight, that comes in two factors:
  \begin{itemize}
    \item[-] an \alphas\ correction: for each clustering $i$ an \alphas\ correction $\alphas(t_{i})/\alphas^{ME}$ is applied;
    \item[-] a Sudakov form factor correction is applied, to make states exclusive.
  \end{itemize}  
  \item Events are accepted or rejected according to their weight.
  \item The accepted events are showered with a veto on the emission above $y_{cut}$.
\end{enumerate}

The matching prescription implemented in \alpgen\ is known as MLM \cite{mlm}. It is similar in the motivations to the CKKW, but different in the implementation. 
Basically, instead of calculating Sudakov weights analytically, the event weighting is performed numerically.
As in CKKW, ME cross sections are calculated up to the maximum parton multiplicity that the user wants in the final state; a minimum \pt\ cut for final state partons is used to cutoff ME divergences and a fixed \alphas\ is used. 
As in CKKW a ``PS history'' is reconstructed and a splitting sequence is identified, with corresponding scales; an \alphas\ correction is applied as in the CKKW. 
From this point the two prescriptions become different. 
In the MLM approach a conventional PS program is used (\pythia\ or \herwig) and ME partons are showered without any constraint. 
The parton collection that results from the PS step is clustered using a jet algorithm (a cone in the \alpgen\ implementation, but also other options have been investigated, e.g. a \kt\ algorithm is used in \madgraph); the resulting jets are matched in angle to the ME partons and only those events in which all the jets match the ME partons without any extra unmatched jets are retained (for the maximum ME parton multiplicity additional jets, softer than the matched ones, are allowed). 

This procedure tries to reproduce in one go the effect that in the CKKW is achieved in two steps: the Sudakov re-weighting and the vetoed shower.
Indeed the rejection of events with additional jets should, at the same time, reject ME configuration in a similar way as the Sudakov weight does and prevent additional emission from the shower, thus reproducing the effect of the CKKW PS veto.

The MLM prescription is really convenient because it does not require modifications in the PS program. It just requires a veto routine to kill events not fulfilling the matching criteria.

While the CKKW prescription contains one parameter (the \qcut\ of the internal \kt\ clustering algorithm), in the MLM the user has to choose different parameters. 
The cone algorithm used for the matching has three parameters, namely the minimum jet \pt\, the cone radius $R$, and the jet maximum pseudorapidity $\eta$. 
The minimum \pt\ used in the cone clustering (\pt$_{min}^{jet}$) is not the same as the minimum \pt\ used in the ME step to cutoff divergences (\pt$_{min}^{ME}$): usually it is recommended to have \pt$_{min}^{jet}>$\pt$_{min}^{ME}$; 
this is needed because events that are below the cut at the ME level could fall above after the PS. 
For this reason a process dependent tuning for \pt$_{min}^{jet}$ with respect to \pt$_{min}^{ME}$ is needed; for the \zjets\ production \alpgen\ authors recommend to choose the jet finder minimum \pt\ to be $5$ GeV higher than the ME minimum \pt.

\section{Generator configuration and analysis}
Plots shown in this paper were obtained with an analysis coded in the Rivet~\cite{hepforgerivet,Buckley:2008zz} MC Validation/Tuning framework. 
Rivet comes with an interface library, called AGILe, which provides interface to many event generators. 
The user can run the event generator through the AGILe interface and analyse the HepMC event record exploiting Rivet analysis classes. 
While Rivet can be used as an analysis framework for general studies on Monte Carlo event generators, as we did in this paper, its main feature is that it comes as a very straightforward tool to make comparisons to data. 

In order to gain a more detailed understanding of the hard event simulation all the generators used in this work were run switching off the underlying event simulation. The analysis has been carried out at parton level.
Also, the emission of final state photons from the leptons from the $Z$ boson decay has been switched off.
In the analysis, the lepton pair from the $Z$ boson decay was required to have an invariant mass between 66 and 116~GeV.
Jets were reconstructed with the longitudinally invariant $k_{\perp}$ algorithm \cite{Catani:1993hr}, as implemented in the \genname{FastJet}\cite{Cacciari:2005hq} package. The pseudo-radius parameter in the $k_{\perp}$ algorithm was set to 0.4 and the minimum \pt\ for jets was set to 30 GeV. 

\section{Matrix element corrections in \pythia\ and \herwig}
\subsection{$Z$ boson transverse momentum}
\begin{figure}[!h]
  \centering\includegraphics[width=0.7\textwidth]{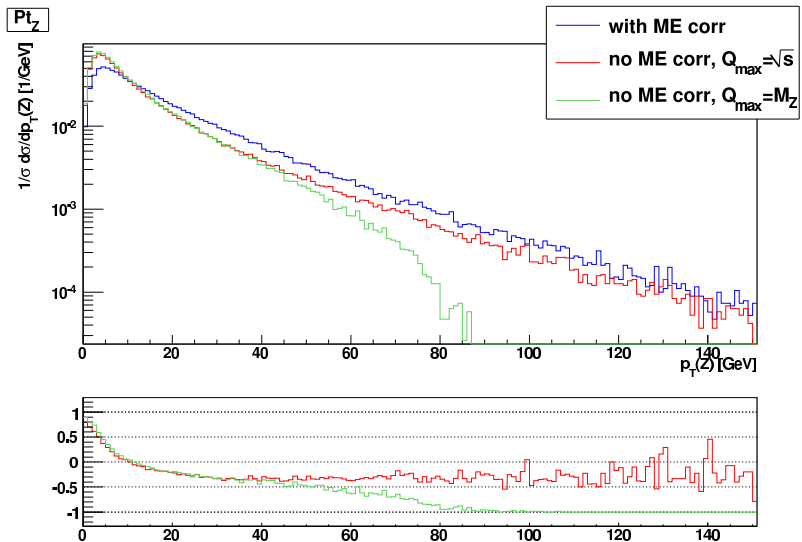}
  \caption{\pt\ spectrum for the lepton pair in \pythia\ for three different settings of the ISR: with ME corrections, without ME corrections and with the starting scale of the shower set to $\sqrt{s}$, without ME corrections and with the starting scale of the shower set to $M_{Z}$. The relative difference with respect to the curve with ME corrections is shown in the lower plot.\label{fig:ptzpythia}}
\end{figure}

The \pt\ distribution for the lepton pair for inclusive $Z/\gamma^{*}$ production in \pythia\ is shown in \Fig{ptzpythia}.
%The lepton pair invariant mass has been generated around the $Z$ resonance, between 66 and 116~GeV. 
Only the electron decay channel has been selected.

The three curves correspond to three different configurations: one is with ME corrections activated, the other two are obtained without matrix element corrections, but with different starting scales for the shower: the total hadronic center of mass energy and the invariant mass of the lepton pair respectively.

\pythia\ implements a virtuality ordered parton shower. The starting scale of the shower marks the maximum allowed virtuality in the shower evolution. 
If the starting scale is set to $M_{Z}$ the hardest parton transverse momentum cannot exceed $M_{Z}$, thus also the $Z$ \pt\ cannot exceed 90 GeV approximately, as shown in \Fig{ptzpythia}. 
When the starting scale is raised, the spectrum gets harder.
When ME corrections are activated the spectrum gets even harder.
The reason why the ME corrected spectrum is harder than the uncorrected one can be explained considering the relative amount of the two corrections at the LHC. The graph $gq\rightarrow{}Z/\gamma^{*}q$ contributes more than $q\bar{q}\rightarrow{}Z/\gamma^{*}g$, because $\bar{q}$ is not a valence quark at LHC. 
%the correction for the $gq\rightarrow{}Z/\gamma^{*}q$ is $>1$ as shown in \Eq{wqgzq}.
We recall that \Eq{wqgzq} states that $W_{gq\rightarrow{}Z/\gamma^{*}q}>1$, meaning that the first emission from the ME-corrected shower is done with a splitting probability higher than that of the uncorrected shower. 
Since the PS emission is always ordered, a higher splitting probability means that the probability for emitting a harder parton is higher than that of uncorrected shower. 
This explains why the corrected spectrum is harder than the uncorrected one.
We also notice in passing that at the Tevatron $p\bar{p}$ collider \cite{bib:tevatron, Lukens:2003aq, Abazov:2002su} exactly the opposite holds: in that case  $q\bar{q}\rightarrow{}Z/\gamma^{*}g$ dominates because $\bar{q}$ is a valence quark; since the correction is $<1$ for this graph (\Eq{wqqbarzg}) the ME corrected result is softer than the uncorrected one, as shown in \cite{Miu:1998ju}.

One might expect that ME corrections should change the shape of spectra only at high \pt; low \pt\ region should be well described by the parton shower alone.
Actually the $Z$ \pt\ spectrum is altered by ME corrections all over the \pt\ range, as shown in \Fig{ptzpythia}.
The three distributions are normalized, but the difference at low \pt\ is not only due to normalization; a change in shape is also present; such a  change is testified  by the relative difference plot, that does not flatten as \pt\ approaches zero.
The reason for the change in shape at low \pt\ is that ME corrections change the Sudakov form factor used in the shower \cite{bib:priv-sjostrand}.

\begin{figure}[!t]
  \centering
  \includegraphics[width=0.7\textwidth]{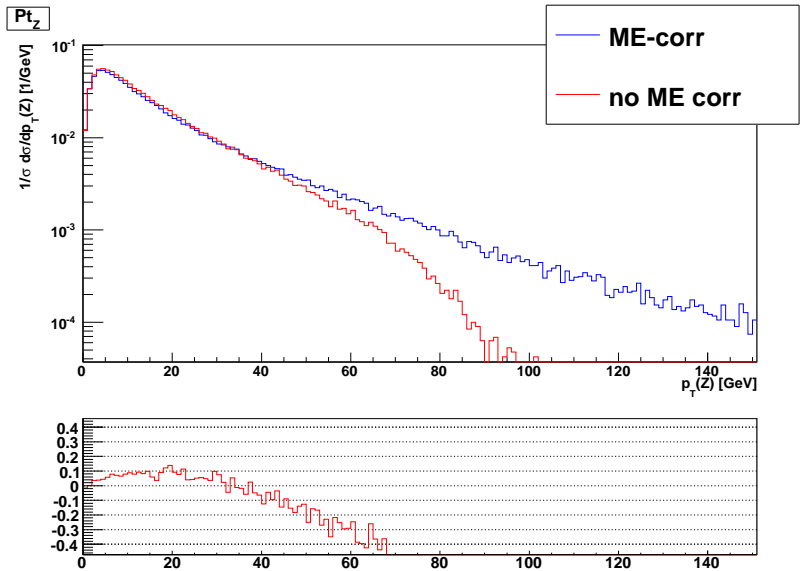}
  \caption{ME corrections effect in \herwig\ in the $Z$ \pt\ distribution. The effect at low \pt\ is small, while in \pythia\ the shape was different at low \pt\ as well.  \label{fig:herwigmecorr}}
\end{figure}

The $Z$ \pt\ spectrum as obtained in \herwig\ is shown in \Fig{herwigmecorr}. Both the ME-corrected and the uncorrected spectra are shown.
While ME corrections in \pythia\ change the whole shape of the distribution, also at low \pt, the change in shape at low \pt\ in \herwig\ is small.

\begin{figure}[!h]
  \centering
  \includegraphics[width=0.7\textwidth]{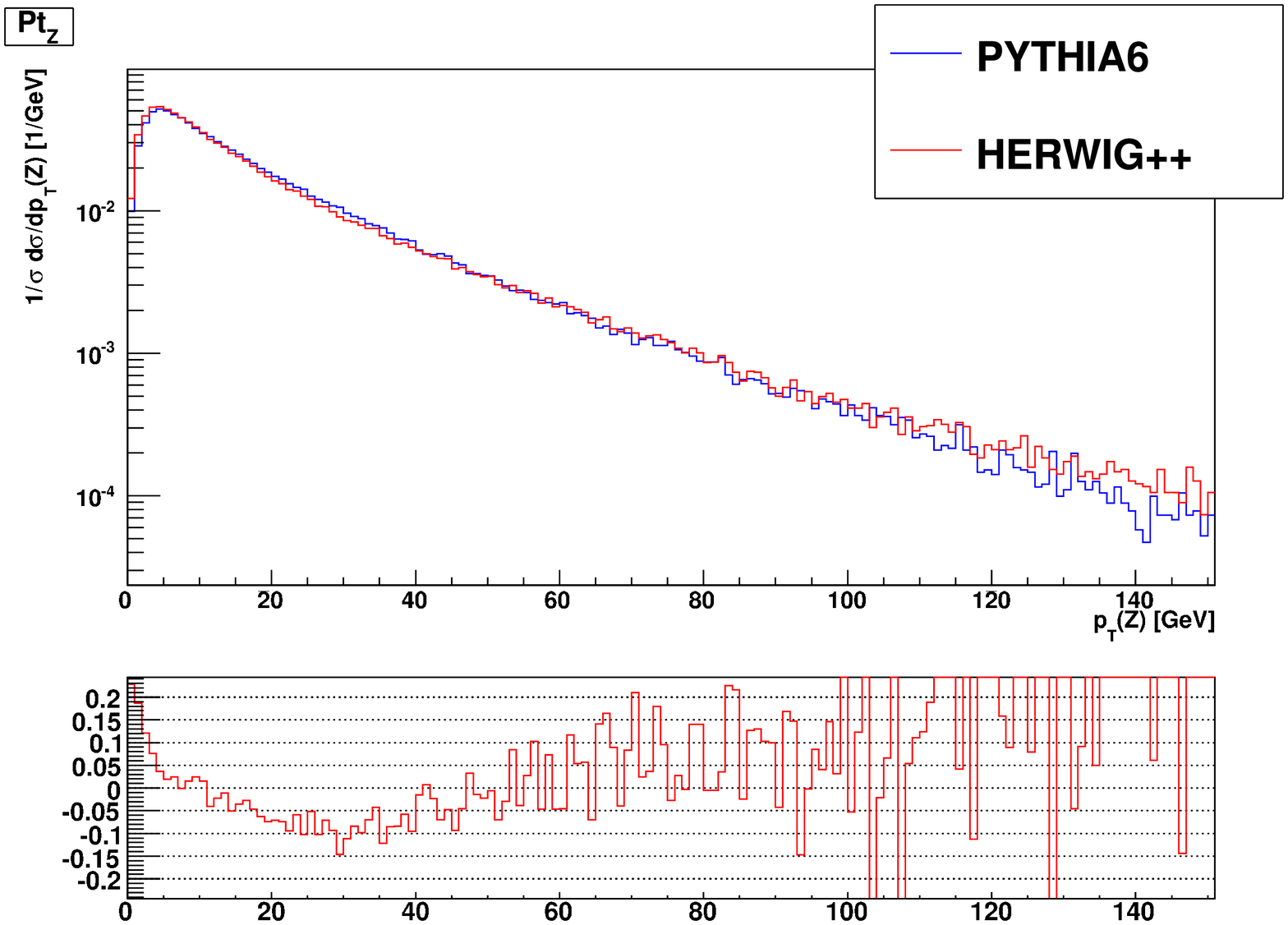}
  \caption{ME corrected Z \pt\ distribution in \pythia\ and \herwig.  \label{fig:herwigpythiaptz}}
\end{figure}

A comparison between ME corrected $Z$ \pt\ distribution in \pythia\ and \herwig\ is shown in \Fig{herwigpythiaptz}.
The agreement is very good all over the \pt\ spectrum.

\subsection{Differential jet rates}
To test how the phase space available for QCD radiation is filled we looked at differential jet rates. 
The differential jet rate $n\rightarrow{}n-1$ is the distribution of the $n\rightarrow{}n-1$ transition value, $Q_{n\rightarrow{}n-1}$, which is the value of the resolution parameter $d_{cut}$ (in an exclusive $k_{\perp}$ algorithm~\cite{Catani:1993hr}) for which an $n$ jet event turns into an $n-1$ jet event.
To understand what a differential jet rate is let's consider a simple example with three particles in the final state.
Let the particles be $1$, $2$, $3$.
Let $d_{i,B}$ be the $k_{\perp}$ distance of particle $i$ from the beam line and $d_{i,j}$ be the $k_{\perp}$ distance between particles $i$ and $j$.
Suppose that the sequence of ordered distances looks like this:
\begin{equation}
  d_{1,B}<d_{2,B}<d_{1,2}<d_{3,B}<d_{1,3}<d_{2,3}.
  \label{eq:sequence1}
\end{equation}
In an exclusive calculation the first jet that would be recombined with the beam line is particle $1$, thus the $3\rightarrow{}2$ transition value is $Q_{3\rightarrow{}2}=d_{1,B}$.
Then particle $2$ is the next one to be recombined with the beam, thus $Q_{2\rightarrow{}1}=d_{2,B}$.
At this point, if the $d_{cut}$ is raised to be at least $d_{2,B}$ only $d_{3,B}
$ survives in \Eq{sequence1}, thus $Q_{1\rightarrow{}0}=d_{3,B}$.

Let's now consider another example sequence:
\begin{equation}
  d_{1,B}<d_{1,3}<d_{1,2}<d_{2,3}<d_{2,B}<d_{3,B}.
  \label{eq:sequence2}
\end{equation}
As before, $Q_{3\rightarrow{}2}=d_{1,B}$. If we raise $d_{cut}$ to at least $d_{1,B}$ particle $1$ gets clustered with the beam and the new sequence looks like
\begin{equation}
  d_{2,3}<d_{2,B}<d_{3,B},
  \label{eq:sequence2a}
\end{equation}
so particles $2$ and $3$ are going to be clustered in the next step. Thus if $d_{cut}$ is set to be at least equal to $d_{2,3}$ particles $2$ and $3$ are clustered, thus passing from $2$ to $1$ jet. Thus $Q_{2\rightarrow{}1}=d_{2,3}$
After they are clustered the sequence will have one only element, namely $d_{(2,3),B}$; thus $Q_{1\rightarrow{}0}=d_{(2,3),B}$.

\begin{figure}[!h]
  \centering\includegraphics[width=0.7\textwidth]{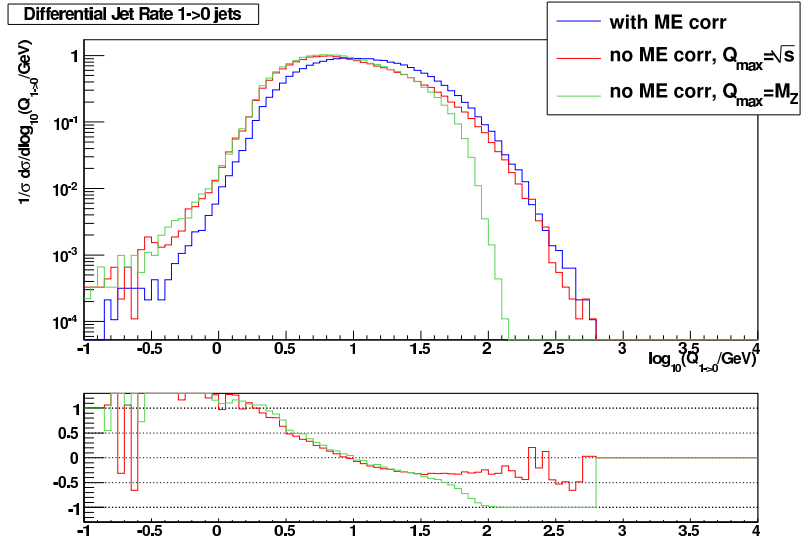}
  \caption{Distribution for the $1\rightarrow{}0$ differential jet rate in \pythia.\label{fig:pythiarate10}}
\end{figure}

Let's now see how the differential jet rates in \pythia\ look for the three setting of ME corrections.
The $1\rightarrow{}0$ differential jet rate is shown in \Fig{pythiarate10}.
The main differences among the three settings are in the region for high values of $Q_{1\rightarrow{}0}$. We see that the 
sample without ME corrections and with low starting shower scale is the one that dies first. This means that it is unable to
 fill the phase space for hard parton emission, which is responsible for filling the rightmost part of the plot.
Regarding the other two settings we observe that they are similar, but while the sample without ME corrections tends to fill the region below 1, the one with ME
 corrections fills the region above 1 more, thus allowing for more radiation to be emitted. 

%Differential jet rates for the transition $2\rightarrow{}1$ and $3\rightarrow{}2$ are shown in \Fig{pythiarate21-32}.
%In this case the differences among the three settings are tamed due to the fact that additional emission other than the first one is anyway uncorrected for ME effects.
%\begin{figure}[!h]
%  \centering
%  \subfigure[]{\includegraphics[width=0.6\textwidth]{pythiarate21.eps}}
%  \subfigure[]{\includegraphics[width=0.6\textwidth]{pythiarate32.eps}}
%  \caption{Differential jet rates for the transitions $2\rightarrow{}1$ (a) and $3\rightarrow{}2$ (b). The effect of different ME correction settings is shown.\label{fig:pythiarate21-32}}
%\end{figure}

\section{\sherpa\ and \alpgen\ compared to \pythia}
As mentioned in the introduction, we made a consistency check of the matching prescriptions implemented in \sherpa\ and \alpgen. 
The test consists in the comparisons of the observables obtained with ME corrected \pythia\ with those obtained with \alpgen\ and \sherpa\ when at most one additional parton is allowed to come from the ME calculation.
This approach has been already explored in \cite{Lavesson:2007uu} to test various matching prescriptions in $e^{+}e^{-}$ collisions.

\subsection{\sherpa}
%We first start considering leptonic observables in \sherpa.
The total cross section for inclusive $Z/\gamma^{*}$ is calculated in \pythia\ at Leading Order accuracy, i.e. it is calculated from the process $q\bar{q}\rightarrow{}e^{+}e^{-}$; the subsequent PS, either corrected or uncorrected, happens with probability 1, so it cannot modify the cross section. 
In \sherpa\ the cross section comes as a sum over the different selected final state parton multiplicities.
\begin{equation}
  \sigma_{\sherpa}=\sum_{i=0}^{N}\sigma_{i}<w_{i}>
  \label{eq:xsec_sherpa}
\end{equation}
where $\sigma_{i}$ is the cross section for $i$ additional partons in the final state and $<w_{i}>$ is the average Sudakov weight for that configuration.

\Tab{xsec_sherpa_pythia} shows the cross section values as obtained in \pythia\ and in \sherpa\ for four different values of the resolution cut \qcut\ that steers the separation between the ME and the PS regions.
\begin{table}[!h]
  \centering  
  \begin{tabular}{ccccc}
    %sherpa              xsec            weight          xsection
    %------------------------------------------------------------
    %Qcut10  |  0 jet                               |      
    %          -------------------------------------
    %        |  1 jet                               |
    %------------------------------------------------------------          
    %Qcut20  |  0 jet                               |     
    %          -------------------------------------
    %        |  1 jet                               |
    %------------------------------------------------------------        
    %Qcut40  |  0 jet                               |
    %          -------------------------------------
    %        |  1 jet                               |
    %------------------------------------------------------------
    %------------------------------------------------------------
    %pythia                                             xsection
    %------------------------------------------------------------
    \sherpa\ &   &          $\sigma_{i}$ [pb] &       $<w_{i}>$ &     Total $\sigma$ [pb]\\
    \hline
    \qcut=10GeV &  0 jet & 838.9 &                0.7489 &    1383\\                               
    &                  1 jet & 998.7 &                0.7559 \\   
    \hline
    \qcut=20GeV &  0 jet & 1059.5 &               0.9301 &    1405\\
    &                  1 jet & 484.6 &                0.8657 \\
    \hline
    \qcut=40GeV &  0 jet & 1271.2 &               0.9992 &    1434\\
    &                  1 jet & 177.2 &                0.9267 \\
    \hline
    \qcut=500GeV &  0 jet & 1926.6 &               0.7540 &   1453\\
    &                   1 jet & 0.038 &                0.9802 \\
    \hline
    \hline
    \pythia\ &   &          &       &     Total $\sigma$ [pb]\\
    inclusive&   &          &       &     1528\\
    \hline
  \end{tabular}
  \caption{Cross sections for \sherpa\ and \pythia.\label{tab:xsec_sherpa_pythia}}
\end{table}

The difference in the total cross section with respect to \pythia\ is up to about 10\%, for the sample with the lowest value of \qcut.
Cross sections for both \sherpa\ and \pythia\ are formally LO; some differences are due to the \qcut\ dependency mainly.
If one takes a very high value for \qcut, this makes \sherpa\ get closer and closer to \pythia.
In fact, as \qcut\ is increased the contribution to the total cross section from the configuration with one additional parton vanishes, thus leaving the leading order contribution alone, which is the only one considered in \pythia\ for the cross section calculation.
In summary the dependency on \qcut\ in \sherpa\ is of the order of 5-10\%. Even when higher order emissions are completely removed there is a 5\% discrepancy with \pythia. This presumably happens due to differences in the choice of scales in the two generators. In particular in \sherpa\ the renormalization scale is set to the \qcut\ value.
%In the sample with \qcut=500~GeV the difference with respect to \pythia\ goes down to 5\%. This residual difference is motivated by different scale choices in the two generators. In \sherpa\ the renormalization scale is set to the \qcut\ value.

\begin{figure}[!t]
  \centering
  \subfigure[]{\includegraphics[width=0.6\textwidth]{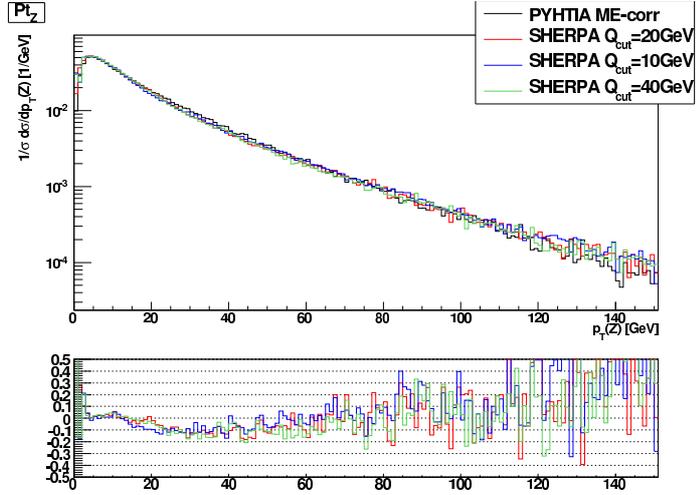}}
  \subfigure[]{\includegraphics[width=0.6\textwidth]{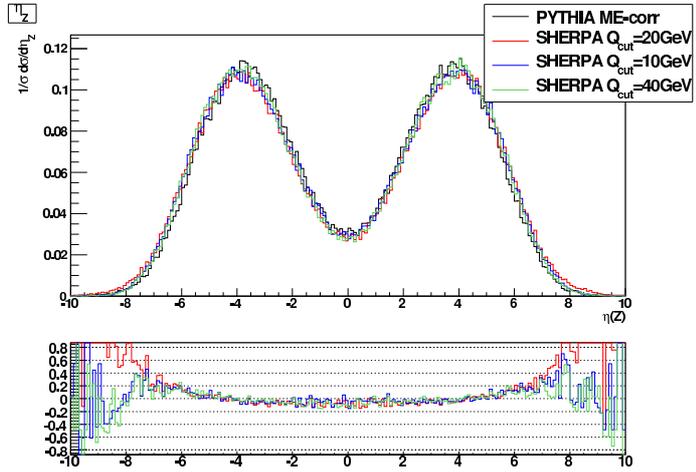}}
  \caption{\pt\ spectrum (a) and $\eta$ distribution (b) for the lepton pair in \pythia\ and \sherpa. The latter has been run with at most one additional parton from the ME; three different values for the separation cut between ME and PS regions have been used: \qcut= 10, 20, 40~GeV.\label{fig:ptetazsherpapythia}}
\end{figure}

The \pt\ spectrum for the lepton pair in \sherpa\ and \pythia\ is shown in \Fig{ptetazsherpapythia} (a). As mentioned above, \sherpa\ has been run such that only one additional parton can be emitted from the matrix element. 
%thus the graphs that contribute to the final state in \sherpa\ are exactly the same \pythia\ is corrected for.
\begin{figure}[!h]
  \centering
  \centering\includegraphics[width=0.6\textwidth]{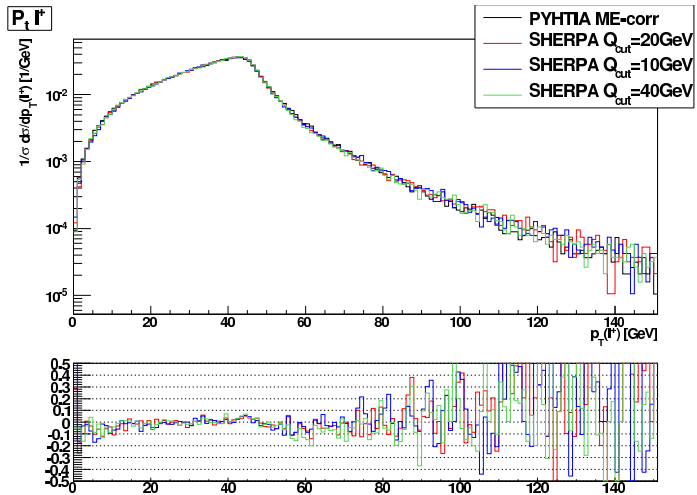}
  %\subfigure[]{\centering\includegraphics[width=0.6\textwidth]{sherpa-pythia/ptl-}}
  \caption{\pt\ distributions for the positive lepton.\label{fig:ptlepsherpapythia}}
\end{figure}

\sherpa\ has been run with three different values of the parameter \qcut\ that governs the separation between the phase space region filled by the ME and the region filled by the PS.
The values used were \qcut= 10, 20, 40~GeV.
As described in \Sec{ckkwmlm} the region above \qcut\ is filled by a modified ME, in which Sudakov form factors are attached to the ME, while the region below the cut is filled by a vetoed PS.
\sherpa\ appears to closely follow the \pythia\ spectrum. 
Some discrepancies are observed in the first few bins. 
These are most probably due to a different tuning for the primordial \pt\footnote{The primordial \pt{} distribution of partons in protons, often referred to as primordial $k_{\perp}$ is the transverse momentum distribution of partons in the hadrons entering the collision. The primordial \kt\ models a soft non-perturbative effect.} 
distribution of partons in the protons.
We notice in particular that the high \pt\ tail, that is sensitive to additional hard emission, appears to be correctly reproduced. 
We also observe a small dependence on the value of \qcut: the three curves for \sherpa\ agree within few percent.

The pseudorapidity $\eta$ distribution for the lepton pair is shown in \Fig{ptetazsherpapythia} (b).
The distributions from \sherpa\ well agree with \pythia\ in the central region, where the difference is within 10-15\%; in the tails of the distribution difference is more evident.

The \pt\ distribution for the positive lepton is shown in \Fig{ptlepsherpapythia}.
In this case too \sherpa\ agrees with \pythia, without strong dependency on the resolution cut \qcut.

\begin{figure}[!t]
  \centering\includegraphics[width=0.7\textwidth]{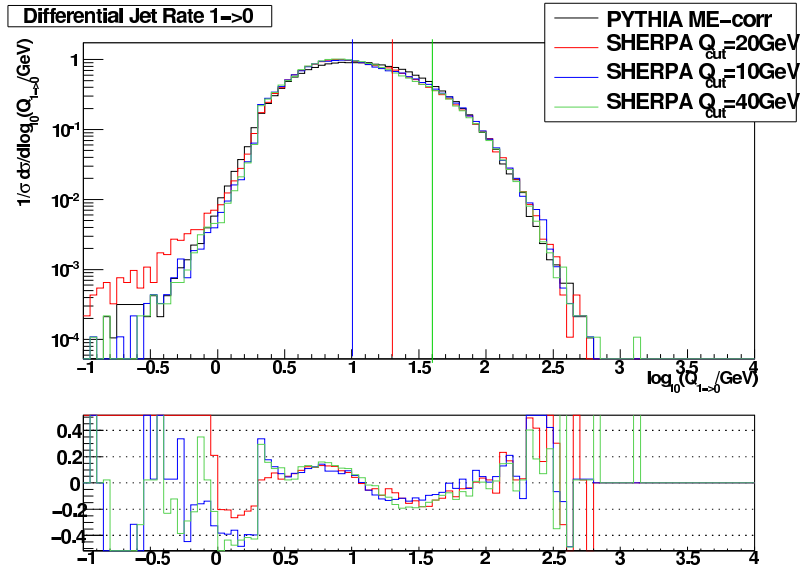}
  \caption{Differential jet rate $1\rightarrow{}0$ in \pythia\ and \sherpa. For \sherpa\ we used three values for the matching parameter \qcut: 10, 20, 40~GeV. Relative differences are calculated with respect to \pythia.\label{fig:rate10sherpapythia}}
\end{figure}

Concerning the QCD observables we looked at differential jet rates.
The distribution for the rate $1\rightarrow{}0$ in \pythia\ and in \sherpa\ is shown in \Fig{rate10sherpapythia}. For \sherpa\ we tried three different values for the matching parameter \qcut: 10, 20, 40~GeV.
The vertical lines shown in the plots indicate the position of the resolution parameter \qcut.
In \sherpa, the phase space above \qcut\ is filled by the Matrix Element, while the region below \qcut\ is filled by the vetoed shower.

\sherpa\ agrees well with \pythia. The transition between the ME- and PS-populated regions is quite smooth for all the three values used for \qcut. 
The relative difference with respect to \pythia\ is at most 20\%, in the vicinities of the \qcut\ region.
The tail for very low values of $Q_{1\rightarrow{}0}$ shows some differences, that can be due to slightly different settings for the primordial \kt. %\pdfmarginnote{di questo sono praticamente sicuro. magari se c'e' tempo giocare con il prim kt per farli uguali}

%For what concerns the rates $2\rightarrow{}1$ and $3\rightarrow{}2$, those are shown in \Fig{rate21-32sherpapythia}.
%The difference with respect to \pythia\ are of the same order or less than in the case of the $1\rightarrow{}0$ transition; apart from a discrepancy in the low $Q$ region, that is presumably due to primordial \kt, the three curves for \sherpa\ agree very well, and the relative difference with respect to \pythia\ is within 15\%.
%This is consistent with what is expected, since the second and third emission are governed by the PS alone both in \pythia\ and \sherpa.
%\begin{figure}[!t]
%  \centering
%  \subfigure[]{\includegraphics[width=0.6\textwidth]{sherpa-pythia_rate21.eps}}
%  \subfigure[]{\includegraphics[width=0.6\textwidth]{sherpa-pythia_rate32.eps}}
%  \caption{Differential jet rate for the transition $2\rightarrow{}1$ (a) and $3\rightarrow{}2$. Relative differences are calculated with respect to \pythia.\label{fig:rate21-32sherpapythia}}
%\end{figure}

\subsection{\alpgen\ plus \pythia}
We made the same test with \alpgen, looking for differences with respect to ME corrected \pythia.
We considered the contribution from up to one additional parton from the matrix element, and we used \pythia\ to shower the ME events generated by \alpgen.

\begin{table}[!b]
  \centering
  \begin{tabular}{cccc}
    \alpgen\ &   &          $\sigma_{i}$ [pb] &     Total $\sigma$ [pb]\\
    \hline
    ME cutoff=10GeV &  0 jet & 1092 &               1609\\
    &                    1 jet & 517\\
    \hline
    ME cutoff=20GeV &  0 jet &  1303&               1594\\
    &                    1 jet &  291 \\
    \hline
    ME cutoff=40GeV &  0 jet & 1452 &               1558\\
    &                    1 jet & 106\\
    \hline
    \hline
    \pythia\ &   &          &       Total $\sigma$ [pb]\\
    inclusive&   &          &       1528\\
    \hline
  \end{tabular}
  \caption{Cross sections for \alpgen\ and \pythia.\label{tab:xsec_alpgen_pythia}}
\end{table}
\Tab{xsec_alpgen_pythia} summarizes the total cross section for \alpgen\ with up to one additional parton from the ME.
Results for three values of the ME cutoff for the generation of the additional jet are shown.
The difference with respect to \pythia\ is 5\% at most.

\begin{figure}[!h]
  \centering
  \subfigure[]{\includegraphics[width=0.6\textwidth]{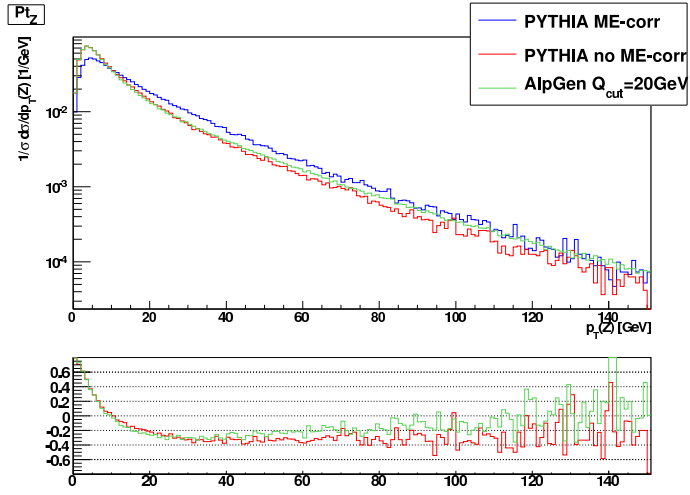}}
  \subfigure[]{\includegraphics[width=0.6\textwidth]{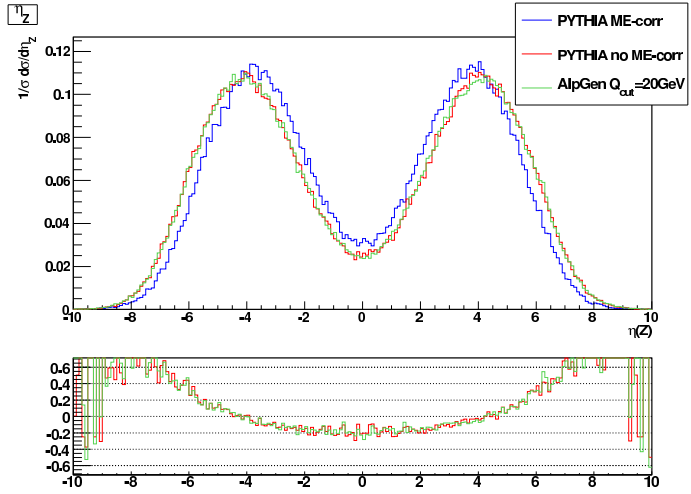}}
  \caption{\pt\ spectrum (a) and $\eta$ distribution (b) for the lepton pair. \alpgen\ is compared with \pythia, with and without ME corrections. Relative difference with respect to ME corrected \pythia\ is shown for each plot.\label{fig:ptzpythiawithwocorralpgen}}
\end{figure}
The \pt\ spectrum and the $\eta$ distribution for \pythia\ (with and without ME corrections) and \alpgen\ are shown in \Fig{ptzpythiawithwocorralpgen}.
Concerning \pythia, the sample without ME corrections has the shower starting scale set to $\sqrt{s}$.
For \alpgen\ we used a minimum \pt\ for the additional ME generated parton of 20~GeV, and the minimum \pt\ for the cone algorithm that steers the matching was set to 25~GeV, as recommended by the authors.
%A detailed explanation of why the matching scale needs to be slightly higher than the ME cutoff in MLM is given in \cite{lavesson-2008-04}.
%\pdfmarginnote{e' il caso che provi a spiegralo ? anche no?}

\alpgen\ lepton pair \pt\ spectrum appears to be softer than ME-corrected \pythia. 
This also translates into a broader $\eta$ distribution for \alpgen.
If we compare \alpgen\ to both ME corrected and to uncorrected \pythia, it appears that \alpgen\ follows uncorrected \pythia\ for low \pt\ values and then agrees with corrected \pythia\ in the high \pt\ tail.

We think that this effect is not related to \alpgen\ itself but rather to \pythia.
When \pythia\ is used to shower \alpgen\ events \pythia's native ME corrections are switched off, because \alpgen\ is going to supply such corrections. As noticed in \Fig{ptzpythia}, the side effect of switching off ME corrections in \pythia\ is that the low \pt\ shape changes. 
\alpgen\ cannot modify the low \pt\ region, which is completely determined by the PS; it can only add high \pt\ radiation which is actually what it does.

\begin{figure}[!t]
  \centering
  \includegraphics[width=0.5\textwidth]{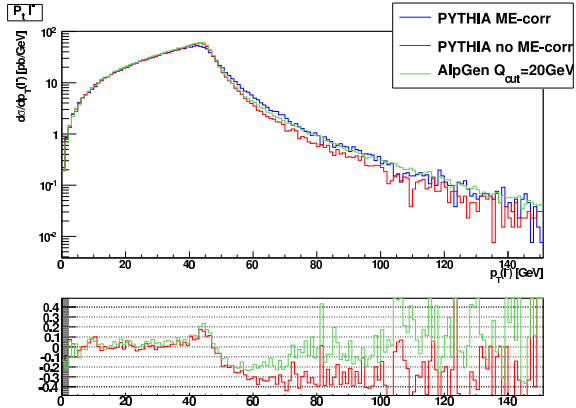}
  %\subfigure[]{\includegraphics[width=0.6\textwidth]{alpgen-pythia/ptl-_qcut20.pdf}}
  %\subfigure[]{\includegraphics[width=0.6\textwidth]{alpgen-pythia/ptl+_qcut20.pdf}}
  \caption{\pt\ spectrum of the electron from $Z$ decay. \alpgen\ is compared with \pythia, with and without ME corrections. Relative differences are with respect to ME-corrected \pythia.\label{fig:ptlpythiawithwocorralpgen}}
\end{figure}

Also the \pt\ spectrum for the leptons from the $Z$ boson decay (\Fig{ptlpythiawithwocorralpgen}) shows the same behavior.
\begin{figure}[!h]
  \centering
  \subfigure[]{\includegraphics[width=0.43\textwidth]{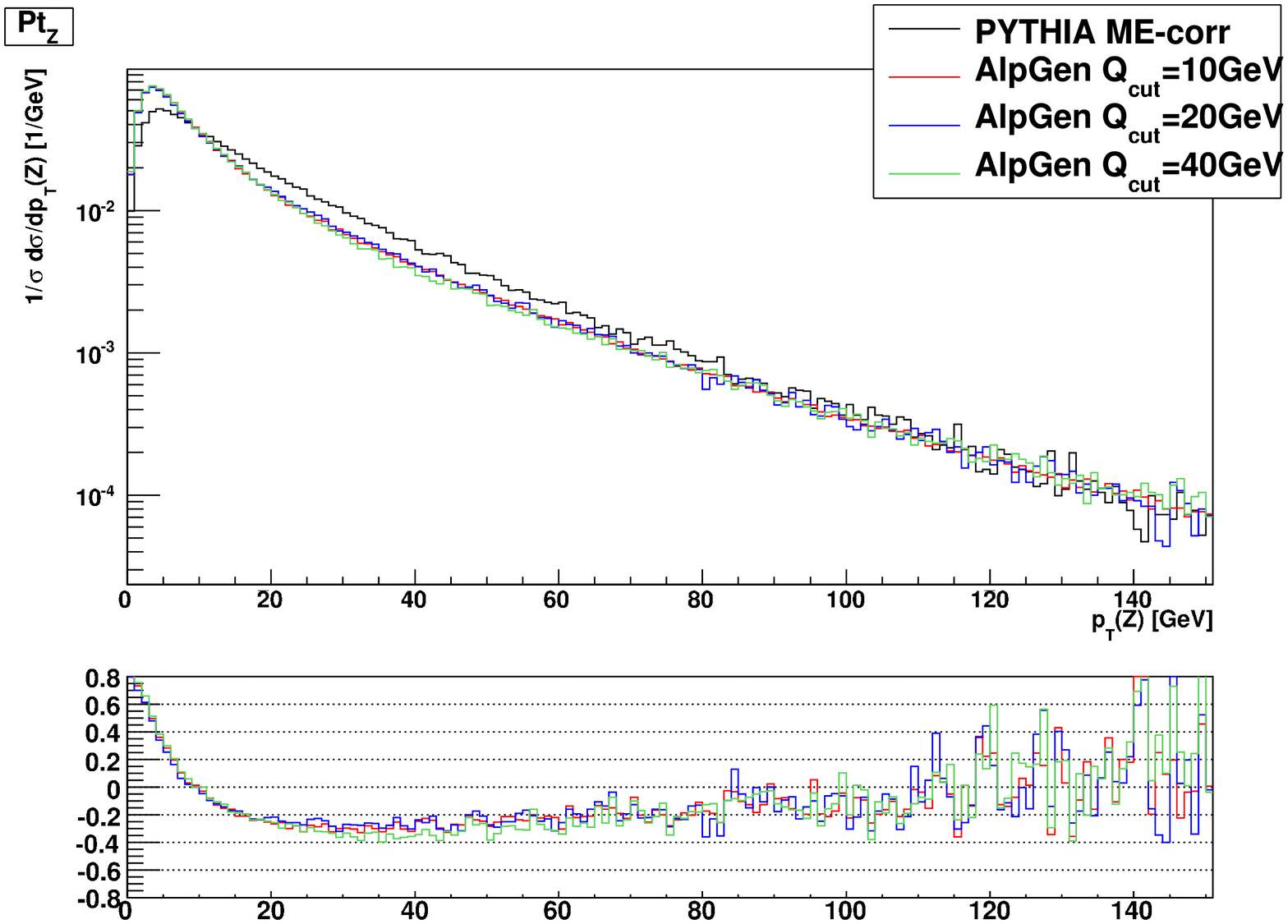}}
  \subfigure[]{\includegraphics[width=0.43\textwidth]{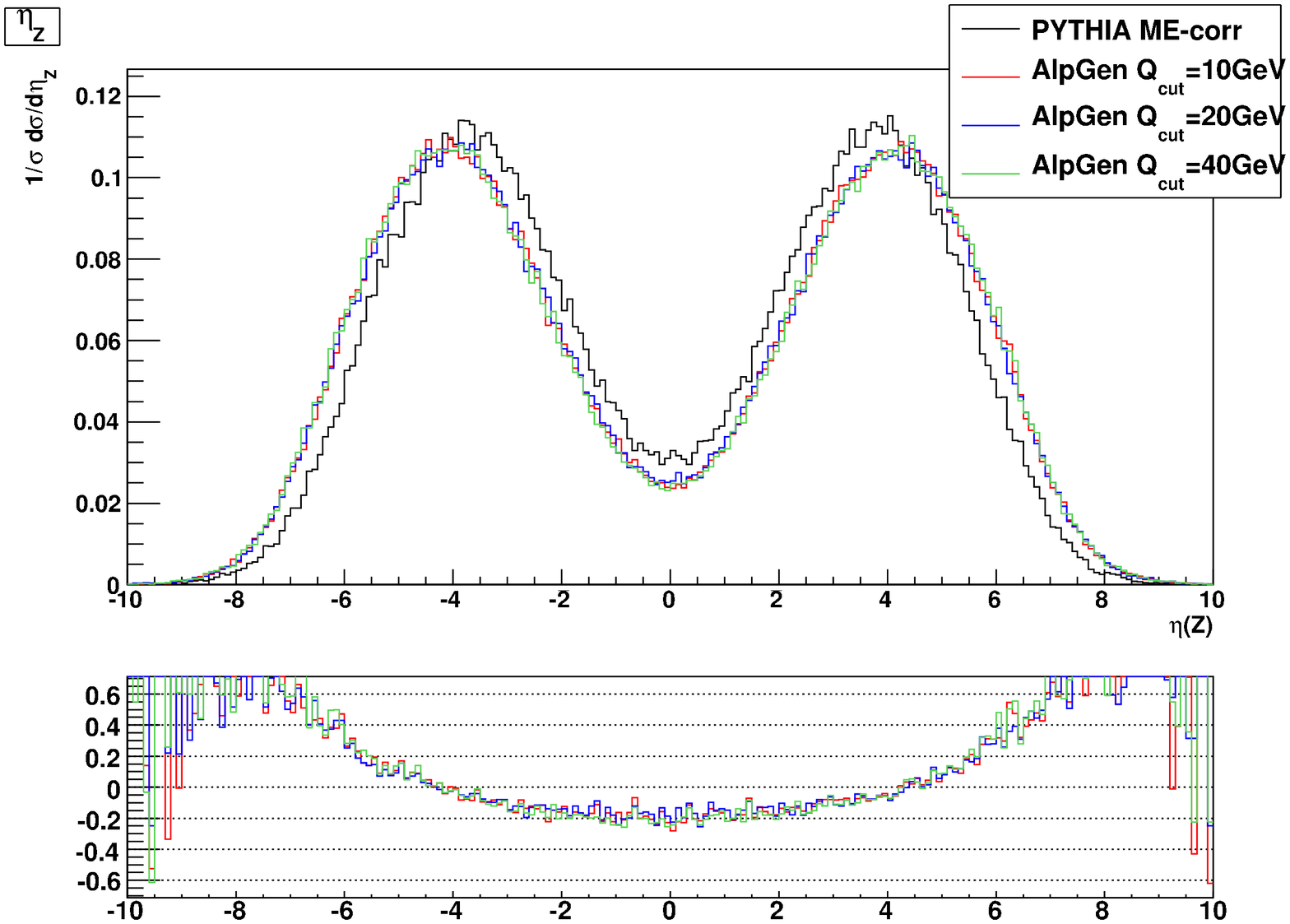}}
  \subfigure[]{\includegraphics[width=0.43\textwidth]{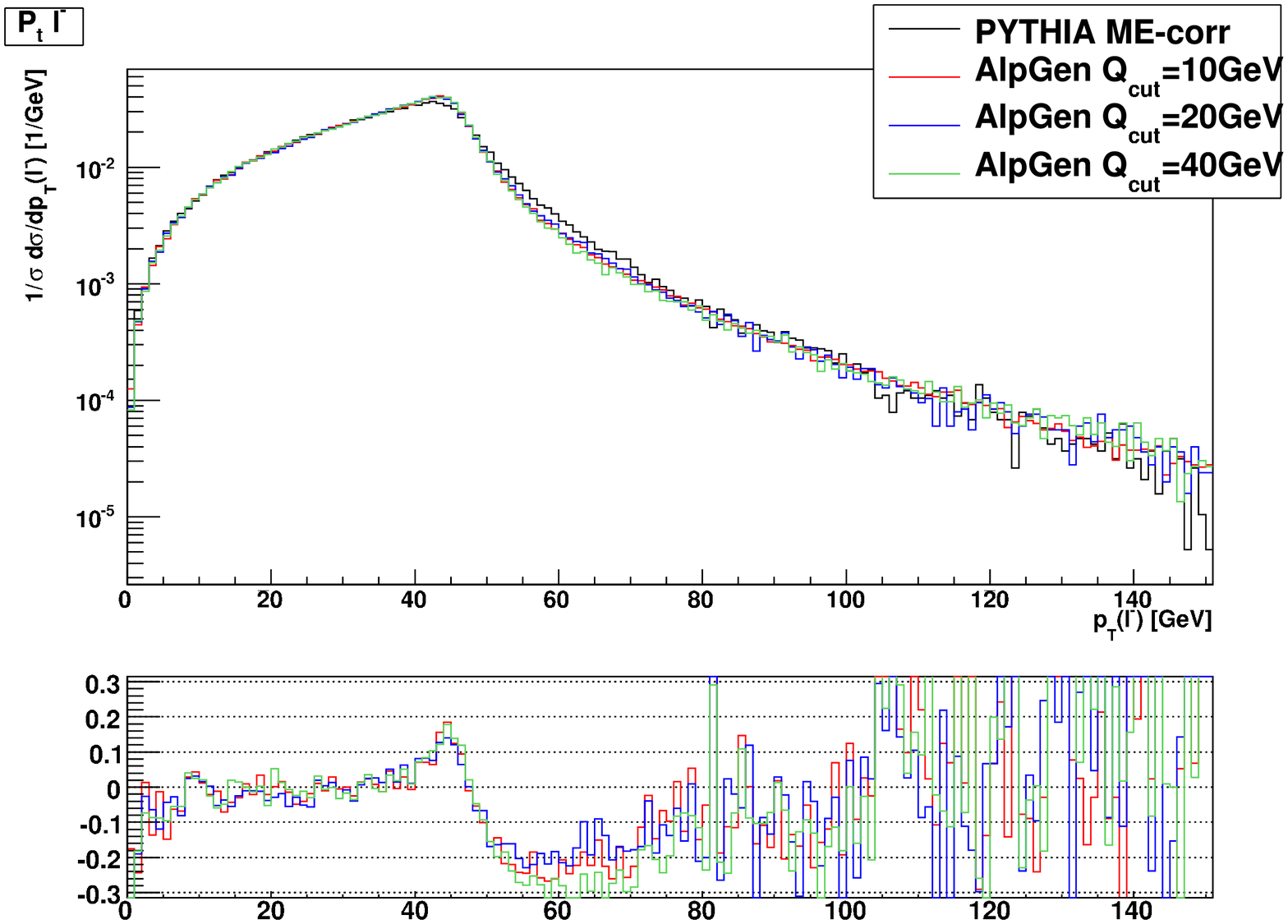}}
  \caption{(a) \pt\ spectrum and (b) $\eta$ distribution for the lepton pair. (c) \pt\ spectrum for the electron. \alpgen\ has been run with three different values of the Matrix Element cutoff. Also ME-corrected \pythia\ is shown as a reference. Relative differences are calculated with respect to ME-corrected \pythia.\label{fig:leptonspythiaalpgenall}}
\end{figure}

The dependency on the resolution cut that separates the ME from the PS region is very limited.
The lepton pair \pt\ and $\eta$ spectra and the \pt\ spectrum for the electron from $Z$ are shown in \Fig{leptonspythiaalpgenall} for three different choices of the Matrix Element cutoff in \alpgen\ (and correspondingly of the minimum \pt\ of the internal cone algorithm): \qcut=10, 20, 40~GeV.

\begin{figure}[!h]
  \centering
  \includegraphics[width=0.6\textwidth]{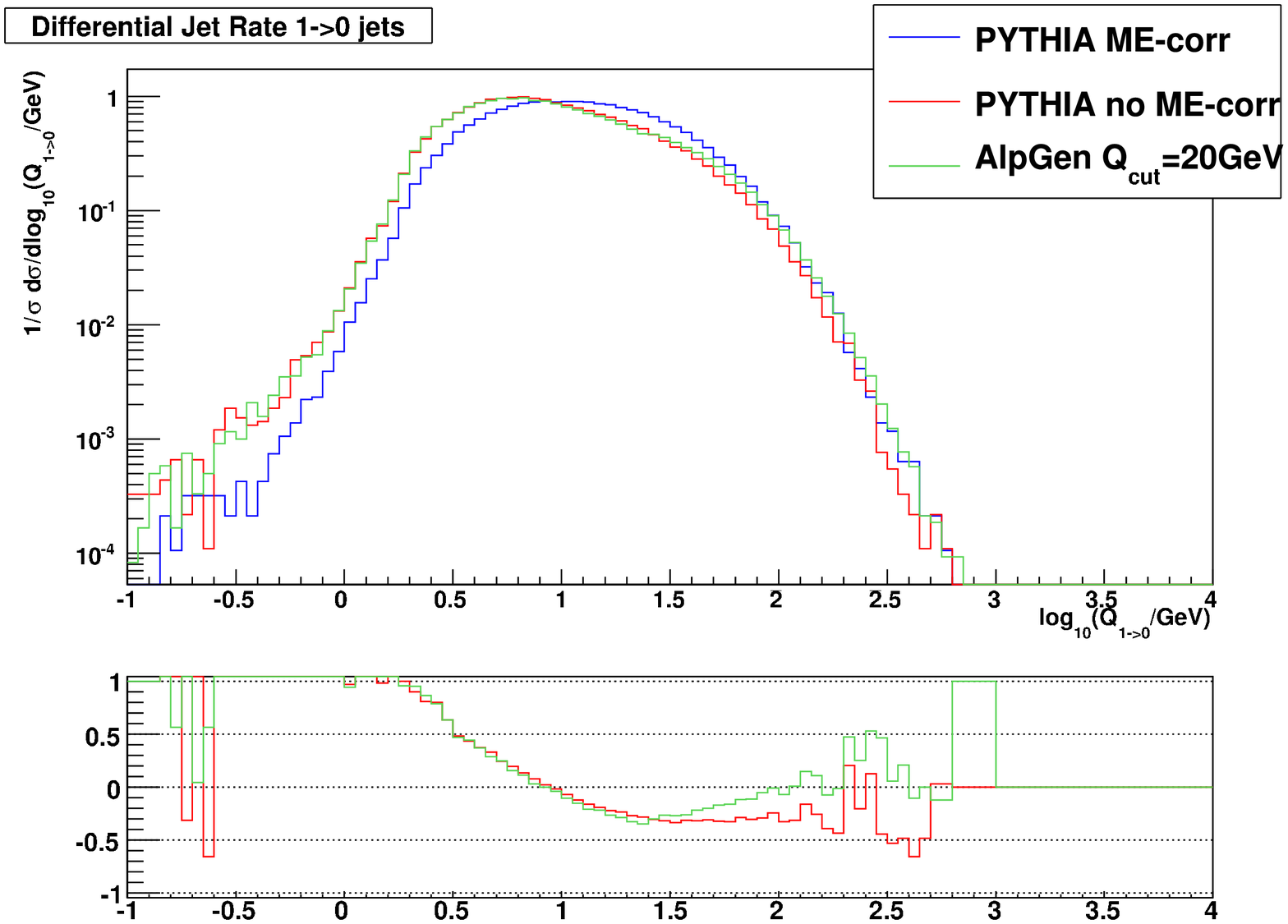}
  \caption{Differential jet rate for the $1\rightarrow{}0$ transition. \alpgen\ has been run with a ME cutoff of 20~GeV. Both ME corrected and uncorrected (starting shower scale set to $\sqrt{s}$) \pythia\ are shown as a reference. Relative differences are with respect to ME corrected \pythia.\label{fig:rate10pythiawithwocorralpgen}}
\end{figure}

Concerning QCD observables, differential jet rate plots appear to confirm the trend observed for letponic observables.
\Fig{rate10pythiawithwocorralpgen} shows the $1\rightarrow{}0$ differential jet rate. Both ME corrected and uncorrected \pythia\ are shown as a reference.
We see that \alpgen\ closely follows uncorrected \pythia\ in the low $Q_{1\rightarrow{}0}$ region, then it starts to agree with corrected \pythia\ for high values of $Q_{1\rightarrow{}0}$.

\begin{figure}[!h]
  \centering
  \includegraphics[width=0.6\textwidth]{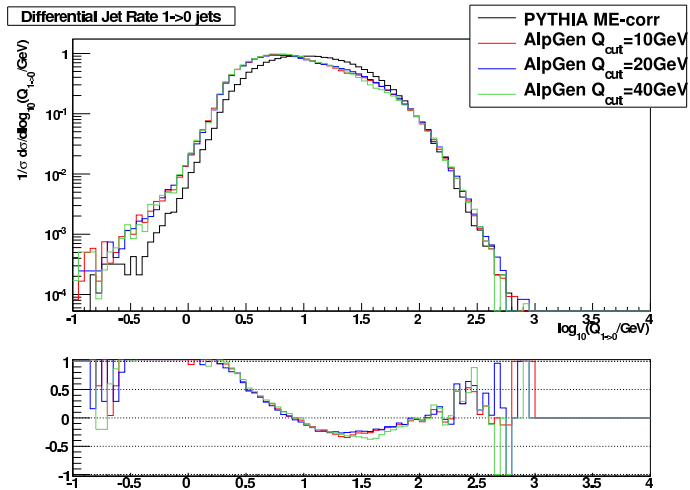}
  %\subfigure[]{\includegraphics[width=0.6\textwidth]{alpgen-pythia_rate10_qall.eps}}
  %\subfigure[]{\includegraphics[width=0.6\textwidth]{alpgen-pythia_rate21_qall.eps}}
  %\subfigure[]{\includegraphics[width=0.6\textwidth]{alpgen-pythia_rate32_qall.eps}}
  %\caption{Differential jet rates for the transitions $1\rightarrow{}0$ (a), $2\rightarrow{}1$ (b) and $3\rightarrow{}2$ (c). \alpgen\ has been run with three different values of the Matrix Element cutoff. Also ME-corrected \pythia\ is shown as a reference. Relative differences are with respect to ME corrected \pythia.\label{fig:ratespythiaalpgenall}}
  \caption{Differential jet rates for the transitions $1\rightarrow{}0$. \alpgen\ has been run with three different values of the Matrix Element cutoff. Also ME-corrected \pythia\ is shown as a reference. Relative differences are with respect to ME corrected \pythia.\label{fig:ratespythiaalpgenall}}
\end{figure}

%\Fig{ratespythiaalpgenall} shows the differential jet rates $1\rightarrow{}0$, $2\rightarrow{}1$ and $3\rightarrow{}2$ in \alpgen\ for three different values of the ME cutoff in \alpgen. Also ME-corrected \pythia\ is shown as a reference.
\Fig{ratespythiaalpgenall} shows the  $1\rightarrow{}0$ differential jet rate in \alpgen\ for three different values of the ME cutoff in \alpgen. Also ME-corrected \pythia\ is shown as a reference.
The dependency on the cut is very limited, as already observed for the leptonic observables.
\newpage

\subsection{\alpgen\ plus \herwig}
\begin{figure}[!h]
  \centering\includegraphics[width=0.7\textwidth]{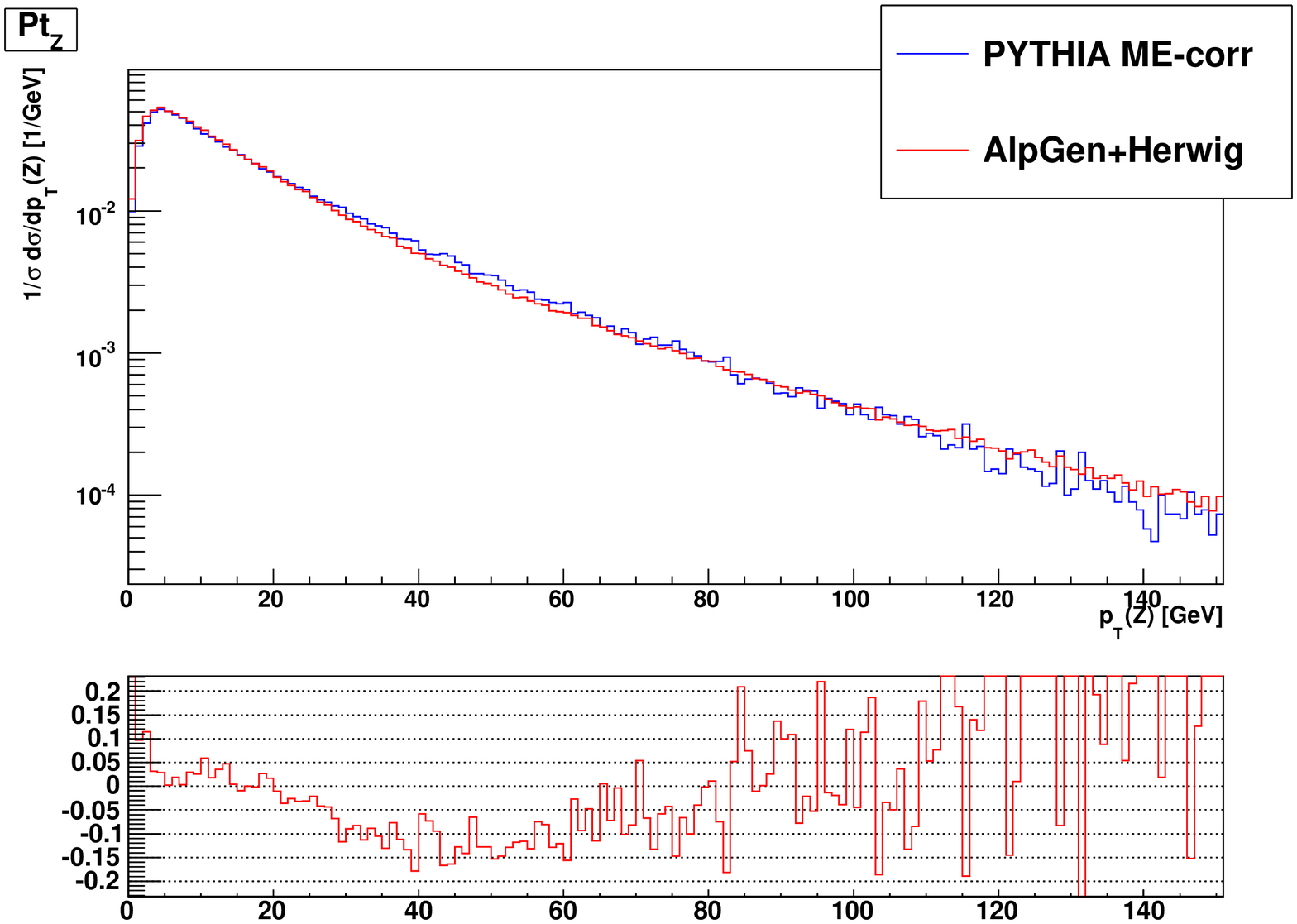}
  \caption{$Z$ boson transverse momentum as obtained in \alpgen+\herwig\ and in \pythia.\label{fig:ptzalpherw}}
\end{figure}

We tried to shower \alpgen\ events also with \herwig. We observed that using \herwig\ gives results much more consistent with the \pythia\ benchmark. 
The $Z$ boson transverse momentum as obtained in \alpgen+\herwig\ is shown in \Fig{ptzalpherw}. ME corrected \pythia\ spectrum is reported in the same plot as a reference.
The agreement is much better than it was using \pythia\ as a parton shower (\Fig{ptzpythiawithwocorralpgen} (a)). The low \pt\ region, that was not reproduced in \alpgen+\pythia\,  is now well reproduced within few percent, except for the very first bins that may be affected by different primordial $k_{\perp}$ tunings.

\section{Conclusion}
We studied the effect of ME corrections in \pythia\ and \herwig.
Both programs can fully take into account ME corrections for one additional parton emission.
The implementation is slightly different.
\begin{itemize}
  \item \pythia\ modifies the shower in two steps: first the starting scale is raised so that any hard emission from the shower is kinematically possible; then the emission probability for the first emission is modified to include ME correction effect.  
  \item In \herwig\ the shower leaves the phase space for hard emission completely uncovered. For this reason the correction is performed in two steps: in the region already filled by the shower the same approach as \pythia\ is used. In the remaining part of the phase space the ME for one additional parton emission is directly used. 
\end{itemize}  
Both approaches give similar results. When ME corrections are switched off the $Z$ \pt\ spectrum in \pythia\ changes also at low \pt, while the low \pt\ shape in \herwig\ remains unchanged.

We used ME corrected \pythia\ as a test bed for \sherpa\ and \alpgen. When those programs are allowed to emit at most one parton from the ME calculation they should give results similar to \pythia.
Actually \sherpa\ follows \pythia\ quite well, both on lepton and jet observables.
On the other hand \alpgen+\pythia\ appears to follow uncorrected \pythia\ at low \pt, for example it shows softer $Z$ \pt\ spectrum with respect to ME corrected \pythia.
This can be traced down to the fact that when \pythia\ is used to shower events produced by \alpgen\ ME corrections are switched off. This is done because \alpgen\ is going to introduce its own corrections.
The side effect of this is that the low \pt\ shape of the $Z$ \pt\ spectrum changes, and \alpgen\ cannot do anything in that region, which is entirely determined by the PS alone. \alpgen\ can only modify the high \pt\ tail of the distribution.

Using \herwig\ to shower \alpgen\ events turns out to be in much better agreement with the \pythia\ benchmark. When using \herwig\ to shower \alpgen\ events native \herwig\ ME corrections are switched off (like in \pythia) but this does not affect the low \pt\ shape, which remains correct.  

The dependency on the cut used to separate the ME and PS regions is limited both in \alpgen\ and in \sherpa.

\appendix
\section{Generator versions and parameters}
In this work we used \pythia\ version 6.411, \sherpa\ version 1.1.2, \herwig++ version 2.2.1. \alpgen\ version 2.13 was used together with \pythia\ 6.411 and \herwig\ 6.510. 

We used \genname{CTEQ6L}\cite{Tung:2002vr} parton density functions.

The simulation of multiple interactions has been switched off for all the generators; also the QED radiation off final state leptons has been switched off.

Different settings for ME corrections in \pythia\ presented were obtained using parameter \genname{MSTP(68)}. In \herwig++ we used parameter \genname{Evolver:MECorrMode} to switch on and off ME corrections.

\section*{Acknowledgements}
We would like to thank Prof. F.~Krauss, Dr. F.~Piccinini, Prof. S.~Catani for many useful discussions. 

This work was supported by the Marie Curie Research Training Network ``MCnet'' (contract number MRTN-CT-2006-035606).

\bibliography{bibliography}

\providecommand{\href}[2]{#2}\begingroup\raggedright\begin{thebibliography}{10}

\bibitem{Sjostrand:2006za}
T.~Sjostrand, S.~Mrenna, and P.~Skands, ``{PYTHIA 6.4 physics and manual},''
  {\em JHEP} {\bf 05} (2006)  026,
\href{http://arxiv.org/abs/hep-ph/0603175}{{\tt arXiv:hep-ph/0603175}}.
%%CITATION = HEP-PH/0603175;%%.

\bibitem{Corcella:2002jc}
G.~Corcella {\em et al.}, ``{HERWIG 6.5 release note},''
\href{http://arxiv.org/abs/hep-ph/0210213}{{\tt arXiv:hep-ph/0210213}}.
%%CITATION = HEP-PH/0210213;%%.

\bibitem{Bahr:2008pv}
M.~Bahr {\em et al.}, ``{Herwig++ Physics and Manual},''
  \href{http://dx.doi.org/10.1140/epjc/s10052-008-0798-9}{{\em Eur. Phys. J.}
  {\bf C58} (2008)  639--707},
\href{http://arxiv.org/abs/0803.0883}{{\tt arXiv:0803.0883 [hep-ph]}}.
%%CITATION = 0803.0883;%%.

\bibitem{Mangano:2002ea}
M.~L. Mangano, M.~Moretti, F.~Piccinini, R.~Pittau, and A.~D. Polosa,
  ``{ALPGEN, a generator for hard multiparton processes in hadronic
  collisions},'' {\em JHEP} {\bf 07} (2003)  001,
\href{http://arxiv.org/abs/hep-ph/0206293}{{\tt arXiv:hep-ph/0206293}}.
%%CITATION = HEP-PH/0206293;%%.

\bibitem{Gleisberg:2008ta}
T.~Gleisberg {\em et al.}, ``{Event generation with SHERPA 1.1},''
  \href{http://dx.doi.org/10.1088/1126-6708/2009/02/007}{{\em JHEP} {\bf 02}
  (2009)  007},
\href{http://arxiv.org/abs/0811.4622}{{\tt arXiv:0811.4622 [hep-ph]}}.
%%CITATION = 0811.4622;%%.

\bibitem{Miu:1998ju}
G.~Miu and T.~Sjostrand, ``{W production in an improved parton shower
  approach},'' \href{http://dx.doi.org/10.1016/S0370-2693(99)00068-4}{{\em
  Phys. Lett.} {\bf B449} (1999)  313--320},
\href{http://arxiv.org/abs/hep-ph/9812455}{{\tt arXiv:hep-ph/9812455}}.
%%CITATION = HEP-PH/9812455;%%.

\bibitem{Seymour:1994df}
M.~H. Seymour, ``{Matrix element corrections to parton shower algorithms},''
  \href{http://dx.doi.org/10.1016/0010-4655(95)00064-M}{{\em Comp. Phys.
  Commun.} {\bf 90} (1995)  95--101},
\href{http://arxiv.org/abs/hep-ph/9410414}{{\tt arXiv:hep-ph/9410414}}.
%%CITATION = HEP-PH/9410414;%%.

\bibitem{Corcella:1999gs}
G.~Corcella and M.~H. Seymour, ``{Initial state radiation in simulations of
  vector boson production at hadron colliders},''
  \href{http://dx.doi.org/10.1016/S0550-3213(99)00672-0}{{\em Nucl. Phys.} {\bf
  B565} (2000)  227--244},
\href{http://arxiv.org/abs/hep-ph/9908388}{{\tt arXiv:hep-ph/9908388}}.
%%CITATION = HEP-PH/9908388;%%.

\bibitem{Catani:2001cc}
S.~Catani, F.~Krauss, R.~Kuhn, and B.~R. Webber, ``{QCD matrix elements +
  parton showers},'' {\em JHEP} {\bf 11} (2001)  063,
\href{http://arxiv.org/abs/hep-ph/0109231}{{\tt arXiv:hep-ph/0109231}}.
%%CITATION = HEP-PH/0109231;%%.

\bibitem{Krauss:2002up}
F.~Krauss, ``{Matrix elements and parton showers in hadronic interactions},''
  {\em JHEP} {\bf 08} (2002)  015,
\href{http://arxiv.org/abs/hep-ph/0205283}{{\tt arXiv:hep-ph/0205283}}.
%%CITATION = HEP-PH/0205283;%%.

\bibitem{mlm}
M.~Mangano, ``{The so-called MLM prescription for ME/PS matching},'' October,
  4, 2002.
\newblock Talk presented at the Fermilab ME/MC Tuning Workshop.

\bibitem{hepforgerivet}
``{Rivet web page},'' 2008.
\newblock \url{http://projects.hepforge.org/rivet/}.

\bibitem{Buckley:2008zz}
A.~Buckley {\em et al.}, ``{CEDAR: Progress and status report},''
\href{http://dx.doi.org/10.1088/1742-6596/119/5/052006}{{\em J. Phys. Conf.
  Ser.} {\bf 119} (2008)  052006}.
%%CITATION = 00462,119,052006;%%.

\bibitem{Catani:1993hr}
S.~Catani, Y.~L. Dokshitzer, M.~H. Seymour, and B.~R. Webber, ``{Longitudinally
  invariant K(t) clustering algorithms for hadron hadron collisions},''
\href{http://dx.doi.org/10.1016/0550-3213(93)90166-M}{{\em Nucl. Phys.} {\bf
  B406} (1993)  187--224}.
%%CITATION = NUPHA,B406,187;%%.

\bibitem{Cacciari:2005hq}
M.~Cacciari and G.~P. Salam, ``{Dispelling the $N^{3}$ myth for the $k_t$
  jet-finder},'' \href{http://dx.doi.org/10.1016/j.physletb.2006.08.037}{{\em
  Phys. Lett.} {\bf B641} (2006)  57--61},
\href{http://arxiv.org/abs/hep-ph/0512210}{{\tt arXiv:hep-ph/0512210}}.
%%CITATION = HEP-PH/0512210;%%.

\bibitem{bib:tevatron}
``{Design report Tevatron 1 project},''. FERMILAB-DESIGN-1983-01.

\bibitem{Lukens:2003aq}
{\bf CDF IIb} Collaboration, P.~T. Lukens, ``{The CDF IIb detector: Technical
  design report},''. FERMILAB-TM-2198.

\bibitem{Abazov:2002su}
{\bf D0} Collaboration, V.~M. Abazov {\em et al.}, ``{Run IIb upgrade technical
  design report},''. FERMILAB-PUB-02-327-E.

\bibitem{bib:priv-sjostrand}
T.~Sjostrand and S.~Mrenna. Private communication.

\bibitem{Lavesson:2007uu}
N.~Lavesson and L.~Lonnblad, ``{Merging parton showers and matrix elements --
  back to basics},''
  \href{http://dx.doi.org/10.1088/1126-6708/2008/04/085}{{\em JHEP} {\bf 04}
  (2008)  085},
\href{http://arxiv.org/abs/0712.2966}{{\tt arXiv:0712.2966 [hep-ph]}}.
%%CITATION = 0712.2966;%%.

\bibitem{Tung:2002vr}
W.-K. Tung, ``{New generation of parton distributions with uncertainties from
  global QCD analysis},'' {\em Acta Phys. Polon.} {\bf B33} (2002)  2933--2938,
\href{http://arxiv.org/abs/hep-ph/0206114}{{\tt arXiv:hep-ph/0206114}}.
%%CITATION = HEP-PH/0206114;%%.

\end{thebibliography}\endgroup


\providecommand{\href}[2]{#2}\begingroup\raggedright\endgroup
\bibliographystyle{utphys}

\end{document}